\documentclass[usenatbib]{mn2e}
\input{psfig}
\usepackage{natbib}  
\usepackage{varioref}

\begin{document} 
\def\lsim{\mathrel{\hbox{\rlap{\hbox{\lower4pt\hbox{$\sim$}}}\hbox{$<$}}}}
\def\gsim{\mathrel{\hbox{\rlap{\hbox{\lower4pt\hbox{$\sim$}}}\hbox{$>$}}}}
\def\simlt{\mathrel{\rlap{\lower 3pt\hbox{$\sim$}}
        \raise 2.0pt\hbox{$<$}}}
\def\simgt{\mathrel{\rlap{\lower 3pt\hbox{$\sim$}}
        \raise 2.0pt\hbox{$>$}}}

\title[AGN self-regulation in cooling flow clusters]
{AGN self-regulation in cooling flow clusters }
\author[A.~Cattaneo, R.~Teyssier]
{A.~Cattaneo $^{1,2}$, R. Teyssier $^{3,2}$\\
$^1$Astrophysikalisches Institut Potsdam, an der Sternwarte 16, 14482 Potsdam, Germany\\
$^2$Institut d'Astrophysique de Paris, 98bis Boulevard Arago, 75014 Paris, France\\
$^3$CEA Saclay, L'Orme des Merisiers, 91191 Gif-sur-Yvette Cedex, France}

\maketitle 
\begin{abstract}

We use three-dimensional high-resolution adaptive-mesh-refinement simulations  to 
investigate if mechanical feedback from active galactic nucleus jets can halt a massive cooling flow
in a galaxy cluster and give rise to a self-regulated accretion cycle.
We start with a $3\times 10^9M_\odot$ black hole at the centre of a spherical halo with the mass of the Virgo cluster. Initially, all the baryons are in a hot intracluster medium in hydrostatic equilibrium within the dark matter's gravitational potential.
The black hole accretes the surrounding gas at the Bondi rate and a fraction of the accretion power is returned into the intracluster medium mechanically through the production of jets.
The accretion, initially slow ($\sim 2\times 10^{-4}\,M_\odot{\rm\,yr}^{-1}$), 
becomes catastrophic, as the gas cools and condenses in the dark matter's potential.
Therefore, it cannot prevent the cooling catastrophe at the centre of the cluster.
However, after this rapid phase,  where 
the accretion rate reaches a peak of $\sim 0.2\,M_\odot{\rm\,yr}^{-1}$, the cavities inflated by the jets become highly turbulent.
The turbulent mixing of the shock-heated gas with the rest of the intracluster medium
puts a quick end to this short-lived rapid-growth phase.
After dropping by almost two orders of magnitudes, the black hole accretion rate stabilises at $\sim 0.006\,M_\odot{\rm\,yr}^{-1}$, without significant variations for several billions of years, indicating that a self-regulated steady-state has been reached.
This accretion rate corresponds to a negligible increase of the black hole mass over the age of the Universe, but is sufficient to create a
quasi-equilibrium state in the cluster core.

\end{abstract}

\begin{keywords}
{cooling flows ---
galaxies: clusters --
galaxies: active --
galaxies: jets}
\end{keywords}

\section{Introduction}

X-ray observations show that active-galactic-nucleus (AGN) jets 
can inflate large cavities in the hot gas of galaxy clusters 
\citep{arnaud_etal84,carilli_etal94,david_etal01,fabian_etal00,fabian_etal02,mcnamara_etal00,mcnamara_etal01}.
\citet{binney_tabor95}
suggested that this injection of mechanical energy could explain why there is little 
neutral gas and star formation at the centre of galaxy clusters,
despite the fact that the time for radiative cooling is in many cases much shorter than the  Hubble time \citep{fabian94}. 
There are three main alternatives to this proposal.
The first one is that the heating needed to quench a massive cooling flow {\it is} provided by the AGN, 
but radiatively \citep{ciotti_ostriker97,ciotti_ostriker01,sazonov_etal04}, and not mechanically. 
The second is that the cluster core is heated from the outside through
thermal conduction (e.g. \citealp{voigt_fabian04}).
The third is that the gas was preheated prior to the epoch of cluster formation by e.g. kinetic outbursts at the epoch of the formation of elliptical galaxies and that it has slowly been cooling since then \citep{babul_etal02,oh_benson03,mccarthy_etal04}.
However, there are objections to both of the first two alternatives (see the review by \citealp{begelman04}),
while the work of AGN outflows on the intracluster medium (ICM) is observed.

AGN outflows can be enormously powerful \citep{mcnamara_etal05}, but
the question is how effectively this power can be used to heat the ICM in the cluster core, e.g.
if the kinetic energy of the outflow is not converted into thermal energy, a small mass may carry all the energy and escape with it from
the cluster.
Moreover, luminous AGNs are short-lived.
It is not clear that such erratic energy sources will have any long lasting effect. 
The problem is complicated because AGN outflows are non-spherical, time-dependent and turbulent, 
so the only proper method of modelling the interaction between the jets 
and the ICM is numerically.

In the last few years, a number of groups have made an intense effort to simulate the impact of radio jets and bubbles
on the structure of the ICM
\citep{churazov_etal01,quilis_etal01,reynolds_etal01,reynolds_etal02,basson_alexander03,omma_etal04,omma_binney04,ruszkowski_etal04,sijacki_springel06,vernaleo_reynolds06}.
This work has given us a reasonable picture of how the lobes of radio sources are formed and of how they evolve after the central engine has stopped pumping energy into them.
It has helped clarifying the mechanisms through which the mechanical energy can be thermalized, and the issues involved, e.g. the viscosity of the ICM, and 
the difference between heavy and light jets.

In all this studies, the power and the duration of the AGN phase were set by hand. 
The AGN was able to affect the ICM, but the change in the ICM properties was not allowed to have any feedback on the AGN power.
Here, we remove this limitation by considering a model in which: i) the jet power  is proportional to the accretion rate of the central
black hole, and ii) the black hole accretion rate depends on the central density and temperature of the ICM through the \citet{bondi52} model.
Thus, we close the loop and investigate the self-regulation of the AGN - cooling flow system directly.

In this paper we start with a $3\times 10^9M_\odot$ black hole at the centre of a  $1.5\times 10^{14}M_\odot$ halo. We simulate how this system evolves
due to radiative cooling and AGN feedback in order to establish
the pattern of black hole accretion and the capacity of the system to approach a state of equilibrium.

The paper is structured as follows. 
In Section~2 we describe our model for the hydrodynamics of the ICM and its interaction with the central
black hole.
In Section~3 we present the simulations and their results. 
In Section~4 we discuss the conclusions that can be derived from these results. 

\section{The model}

\subsection{The intracluster medium}

We assume that we can treat the ICM as an ideal gas and we follow its dynamics in the gravitational potential of a static spherical dark matter halo, the
radial density profile of which is the described by the NFW model \citep{navarro_etal97}.
We include energy dissipation due to radiative cooling as well as the injection of mass, momentum and energy due to the presence of an accreting black hole at
the centre of the gravitational potential.
The equations of motions in their conservative  form are:
\begin{equation}
\label{continuity}
{\partial\rho\over\partial t}+{\bf\nabla}\cdot{\rho\bf v}=|\psi|\dot{M}_{\rm j}
\end{equation}
\begin{equation}
\label{euler}
{\partial\over\partial t}(\rho{\bf v})+{\bf\nabla}\cdot(\rho{\bf v}\otimes{\bf v})+{\bf\nabla}p=-\rho{\bf\nabla}\phi_{\rm dm}+\psi\dot{p}_{\rm j}{\bf n}_z
\end{equation}
\begin{equation}
\label{energy}
{\partial\over\partial t}(\rho e)+{\bf\nabla}\cdot\left[\rho{\bf v}\left(e+{p\over\rho}\right)\right]=
-\rho{\bf v}\cdot{\bf\nabla}\phi_{\rm dm}-C+|\psi| P_{\rm j}
\end{equation}
\begin{equation}
\label{state}
p=(\gamma-1)\rho\left(e-{1\over 2}{\bf v}^2\right)
\end{equation} 
where $\rho$ is the density, ${\bf v}$ is the velocity, $e$ is the specific
total energy, $p$ is the pressure, $\phi_{\rm dm}$ is the gravitational potential of the dark matter,
$C$ is the radiated power per unit volume), $\gamma$ is the ratio between the specific heats at constant pressure and volume,
$\dot{M}_{\rm jet}$, $\dot{p}_{\rm jet}$ and $P_{\rm jet}$ are the mass, momentum and energy injection rates due to feedback from black hole accretion,
$\psi$ is a function that specifies the spatial distribution of the injection (the integral of $|\psi|$ over space is equal to unity), while ${\bf n}_z$ is the unit
vector in the positive direction of $z$ axis, chosen to coincide with the axis of the jets coming out of the central source.

The radiated power per unit volume has the form $C=n_{\rm H}^2\Lambda(T,Z)$, where
$n_{\rm H}=f_{\rm H}\rho/m_{\rm p}$ is the number density of hydrogen nuclei.
Here $m_{\rm p}$ is the proton's mass and $f_{\rm H}=0.75$ is the mass fraction of the ICM in hydrogen.
The cooling function
$\Lambda$ only depends on the temperature $T$ and the metallicity $Z$.
It is computed from the collisional equilibrium cooling functions tabulated in \citet{sutherland_dopita93}.
The temperature from which we interpolate the cooling rate is computed from $p$ and $\rho$ using the
equation of state for an ideal gas with an  atomic weight
of $\mu\simeq 0.6 {\rm  m_p}$, the value for a totally ionised plasma with $25\%$ helium in mass.
All our calculations are for a metal abundance of one third the Solar value.

In our simulations, we assume that all the baryons are in the ICM and that the cosmic baryonic fraction is $10\%$.
We neglect the fact that the ICM has a non-zero total angular momentum. 
We assume that initially the ICM is in hydrostatic equilibrium within a static NFW potential.
Appendix~A contains the technical details of how we set up our hydrostatic initial conditions.

\subsection{Black hole accretion and feedback}

We start with a supermassive black hole 
at the centre of the gravitational potential.
The black hole growth rate is computed with the Bondi formula.
\citet{bondi52} studied a spherical stationary solution where 
the gas is static and homogeneous at infinity and in free fall close to
the black hole. The transition between the two regimes occurs at the sonic radius $r_{\rm s}$, 
where the circular velocity on a Keplerian orbit around the black hole is equal to the sound speed $c_{\rm s}$ of the ICM 
(${\rm G}M_\bullet/r_{\rm s}=c_{\rm s}^2$).
Under this hypothesis, the black hole accretion rate is
\begin{equation}
\label{bondi_formula}
\dot{M}_\bullet=4\pi r_{\rm s}^2\rho c_{\rm s} =4\pi({\rm G}M_\bullet)^2{\rho\over c_{\rm s}^{3}}
\end{equation}
where $\rho$ is the gas density at the sonic radius.
In reality, the gas distribution around the black hole is non-spherical and turbulent,
so it is difficult to define the sonic radius. 
For this reason, we adopt a practical approach and determine
$\rho$ and $c_{\rm s}$ as the density and the volume-weighted sound speed in a fixed small but resolved volume around the black hole.
It is also worth noting that, for $\gamma=5/3$, $\rho c_{\rm s}^{-3}=(5/3)^{3/2}s^{-3/2}$, where $s\equiv p\rho^{-\gamma}=s_0{\rm\,exp}[(\gamma-1)\mu m_{\rm p}k^{-1}(S-S_0)]$ is a
measure of the entropy per unit mass, $S$, of the ICM ($k$ is the Boltzmann constant and $s_0$ is the value of $s$ that corresponds to the zero
point of the specific entropy, $S_0$) . Hence, in the Bondi model, $\dot{M}_\bullet$
depends on the state of the ICM through its entropy only.

We assume that black hole accretion is accompanied by the acceleration of jets. 
Jets emit radio synchrotron radiation because they are magnetised.
A magnetic field of 1\,nT in vacuum corresponds to
a magnetic pressure of $\sim 10^{-12}{\rm dyn\,cm}^{-2}$, comparable to the thermal pressure of the ICM.
Magnetic fields should therefore be important. They are also believed to play a role in collimating the jet itself.
However, we assume here that we can model jets using only Euler's equations.
This approximation is valid if one is interested in the dynamics of the cocoons inflated by the jets, rather the dynamics of the jets themselves. These hot bubbles are responsible for the mechanical
work on the ICM. As long as the energy and the
momentum pumped into the jets are calibrated correctly, one will end up reproducing the expansion of
the cocoons appropriately.

We assume that the power of the relativistic particles pumped into the jets is proportional to the accretion rate of the black hole
and that on subparsec scales, where jets are accelerated, the outflow rate is equal to the accretion rate and all the energy is kinetic. With these hypotheses,
the rates at which 
momentum and energy are deposited into the jets are:
\begin{eqnarray}
\label{jetspeed}
\dot{p}_{\rm j}  &=&\sqrt{2\epsilon}\,\,\,\,\dot{M}_\bullet{\rm c}\\
P_{\rm j}           &=&\epsilon\,\,\,\,\,\,\,\,\,\,\,\,\dot{M}_\bullet{\rm c}^2
\end{eqnarray}
Here c is the speed of light and $\epsilon$ 
is the efficiency with which the rest-mass energy of matter accreted onto the black hole is used to power the jets. 

As jets travel from the subresolution region where they are accelerated to the smallest scale that we
can resolve in our simulations, they entrain additional material from the
interstellar medium of the black hole's host galaxy. This increases 
the mass outflow rate, while conserving both the momentum and the total energy. Defining  $\eta$ as the jet mass loading factor, we have:
\begin{eqnarray}
\label{outflowrate}
\dot{M}_{\rm j} &=&\eta\,\,\,\,\,\,\,\,\,\,\dot{M}_\bullet
\end{eqnarray}
This process occurs at scales far below our resolution limit.
 We consider here
only the large scale jets, for which the velocity is much smaller than the speed of light.
The jets are launched in two opposite directions from the central $\sim3\,$kpc at a speed
$v_{\rm j}=\dot{p}_{\rm j} /\dot{M}_{\rm j} =\eta^{-1}\sqrt(2\epsilon){\rm c}$. Note that this speed
is consistent with momentum conservation in the jet.
A mass loading factor of $\eta=1$ corresponds to the case without entrainment, in which all the energy is kinetic.
For $\eta>1$, $P_{\rm j}>1/2\dot{M}_{\rm j}v_{\rm j}^2$. The difference between the two is added as thermal energy to the gas injected at the base of the jets. The physical reason is that the entrained gas is shocked. Therefore, kinetic energy is converted into heat.

In our mathematical formulation, $\dot{M}_{\rm j}$, $\dot{p}_{\rm j}$ and $P_{\rm j}$ are function of time only and
as such they do not contain any spatial information, which is, instead, in the function $\psi$ (Eqs.~\ref{continuity},\,\ref{euler},\,\ref{energy}). The precise form of the function $\psi$ is dictated by computational reasons
and has no physical origin.  Its consideration is therefore postponed to the section that deals with the numerical aspects.

\subsection{Choice of physical parameters}

Our choice of environmental and black hole parameters is dictated by observations of M87,
although our study is not an attempt at realistically modelling any individual object.
In particular, we have neglected the stellar potential, which dominates the gravity in the central parts of M87. The halo has a mass of $M_{\rm vir}=1.5\times 10^{14}M_\odot$.
The virial radius, $r_{\rm vir}\simeq 1.4\,$Mpc, is computed by assuming that the virial 
density is $\Delta_{\rm c}=101$ times  the critical cosmic density.
The core radius of the dark matter distribution, $r_0\simeq 250\,$kpc, is computed from the virial radius with the concentration parameter given
by Eq.~(\ref{cnfw}).

The initial mass of the central black hole is $M_\bullet=3\times 10^9M_\odot$. It is commonly assumed that  the accretion of matter onto a black hole releases energy at $\sim 10\%$ efficiency, so that 
the accretion power is $P_{\rm accr}\sim 0.1\dot{M}_\bullet{\rm c}^2$, and that most of this power is emitted as light. 
This is almost certainly true for the AGNs detected in optical surveys.
In fact, there is reasonably good agreement between the cosmic black hole density inferred from
mass estimates in the local Universe
and the density derived by integrating the optical and X-ray emission from AGNs over the entire life of the Universe
\citep{yu_tremaine02,barger_etal05}.

\begin{table}
\caption{Simulation parameters (halo, black hole, numerical)}
\begin{tabular}{l|l}
\hline
\hline
$M_{\rm vir}$    &  $1.5\times 10^{14}M_\odot$\\
$r_{\rm vir}$    &  1.4{\rm\,}Mpc\\
$r_0$    &  250{\rm\,}kpc\\
\hline
$M_\bullet$    &  $3\times 10^{9}M_\odot$\\
$\epsilon$&  0.1\\
$\eta$    &  100\\
\hline
$r_{\rm j}$    &  3.2$\,$kpc\\
$h$    &  2.5$\,$kpc\\
$L_{\rm box}$    &  $648\,$kpc\\
$\Delta r$              &  $0.64\,$kpc\\
$t_{\rm sim}$              &  $12\,$Gyr\\
\hline
\hline
\end{tabular}
\end{table}

However, both observation (M87; \citealp{dimatteo_etal03})
and theory (ADIOS adiabatic inflow-outflow solution; \citealp{blandford_begelman99})
suggest that this may not be true when $\dot{M}_\bullet\ll\dot{M}_{\rm Edd}$, where
$\dot{M}_{\rm Edd}$ is the accretion rate needed for the AGN to radiate at the Eddington luminosity.
In such cases, the most probable outcome is that most of the power is released mechanically.

In the simulation that we present in this article, the maximum accretion rate is 
$\dot{M}_\bullet\sim 0.2\,M_\odot{\rm\,yr}^{-1}$ (Section~3).
For $M_\bullet=3\times 10^9\,M_\odot$, this gives 
$\dot{M}_\bullet\sim 3\times 10^{-3}\dot{M}_{\rm Edd}$.
Therefore, we are at all times inside the regime described by ADIOS,
where we can assume that all the power generated by the AGN is channelled into the jets.
If matter accreted by the black hole releases energy at the canonical efficiency of $\sim 10\%$,
this gives us a canonical value of $\epsilon \sim 0.1$.

We have set the mass loading factor to $\eta=100$
so that  plasma is injected away from them black hole at $v_{\rm j}\sim 1350{\rm\,km\,s}^{-1}$ in agreement
with observations of Fanaroff-Riley I sources.
This speed is much lower than the one at which particles are accelerated close to the central engine because
the jets have entrained mass and have therefore slowed down before reaching the 3{\rm\,kpc} scale
at which we introduce them in our simulations.
The parameter $\eta$ also determines the temperature of the plasma, since the rates at which kinetic energy and thermal energy are produced must add to the total power that the AGN deposits into the ICM. 
Choosing  $v_{\rm j}\sim 1350{\rm\,km\,s}^{-1}$ implies that the kinetic energy is only a small fraction of the total energy of the jets. 
Our jet model is therefore intermediate between those of
\citet{reynolds_etal02} and \citet{omma_etal04} although closer to the former.
This non-relativistic speed also ensures that we are allowed to model the system with classical hydrodynamics.

\subsection{Numerical aspects}

The adaptive mesh refinement (AMR) hydrodynamic code RAMSES \citep{teyssier02}  
is used to integrate the equations of the conservation of mass, momentum and energy 
on a three-dimensional Cartesian grid. 
The size of a computational cell is $2^{-\ell}L_{\rm box}$, where $L_{\rm box}$ is the size of the cubic
computational box and $\ell$ is the level of refinement.
In AMR codes such as RAMSES,
$\ell$ can have a large value where high spatial resolution is needed
(e.g. in the cluster core or along the jet axis) without slowing down the
computation by imposing high resolution over the entire box. 
The time step is also adaptive and is determined for each level of
refinement independently by using standard stability constraints for
hydrodynamic solvers.

Our refinement strategy is based on the geometry of the problem, which is a combination of
the spherical geometry of the potential and the cylindrical geometry of the jets.
The computational domain is a cube of size $L_{\rm box}\simeq 650\,$kpc, 
subdivided in large cells of size  $2^{{-\ell}_{\rm min}}L_{\rm box}$, where ${\ell}_{\rm min}=5$.
We define an ellipsoidal region, centred on the cluster core,
with minor axis $a\simeq 40\,$kpc and major axis $b\simeq160\,$kpc. 
The major axis is aligned with the jet axis.
This region is refined up to the maximum level, $\ell_{max}=10$. 
The cell size at the highest level of refinement is $\Delta r\simeq 0.6$kpc.
We then progressively de-refine the grid outside this central region in order to save memory and computing 
time.  We obtain a total number of AMR cells $N_{tot} \simeq 5 \times 10^5$, so that one run takes 
approximately 100 hours wall clock time on 32 processors for
a total number of time steps roughly equal to $10^6$.

Jet simulations are very time consuming because the time step, controlled
by the Courant stability condition, is very small. This has prevented us from performing
higher resolution simulations to estimate the convergence properties of our settings.
In cooling-flow simulations, the gas temperature is lower and the number of time steps
is one order of magnitude smaller. We have therefore been able to increase the size of the box up to
twice the virial radius and to raise the number of cells up to $5\times 10^6$. 
In this case, we obtain a cooled gas fraction of $\sim 7\%$ instead of the one obtained in our `fiducial' run
(the one with the same resolution and box size that we use for jet simulations),
which is around 3.5\% of the total gas mass.
Due to a limited resolution and box size, we might therefore underestimate
cooling by a factor of $\sim 2$. 
However, due to the self-regulating nature of AGN feedback (Section~3), we are confident that our
ultimate conclusions remain unaffected by these numerical issues.

\begin{figure*}
\noindent
\begin{minipage}{5.6cm}
  \centerline{\hbox{
      \psfig{figure=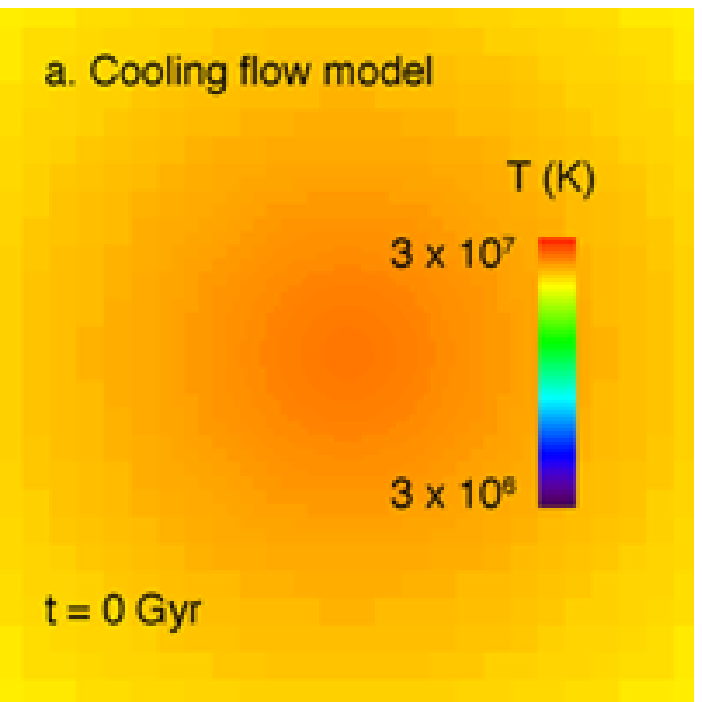,height=5.6cm,angle=0}
  }}
\end{minipage}\    \
\begin{minipage}{5.6cm}
  \centerline{\hbox{
      \psfig{figure=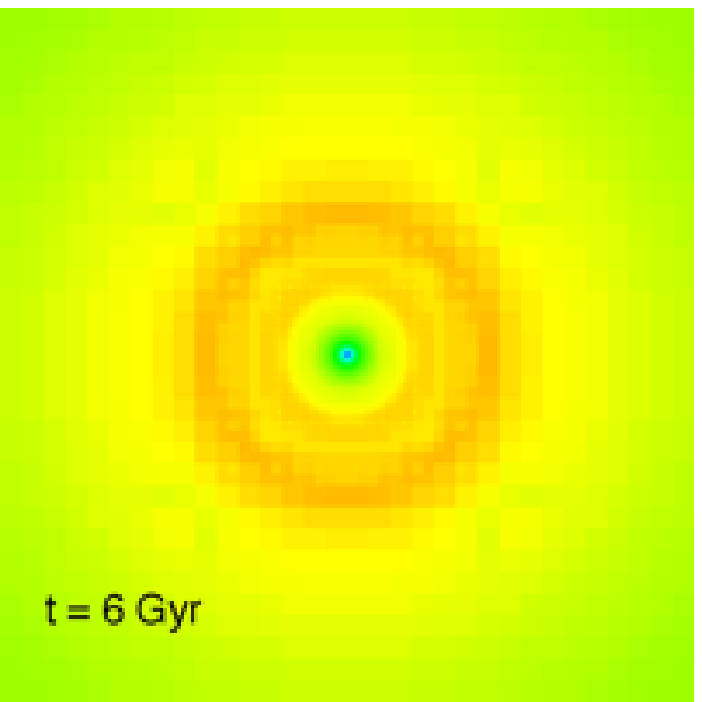,height=5.6cm,angle=0}
  }}
\end{minipage}\    \
\begin{minipage}{5.6cm}
  \centerline{\hbox{
      \psfig{figure=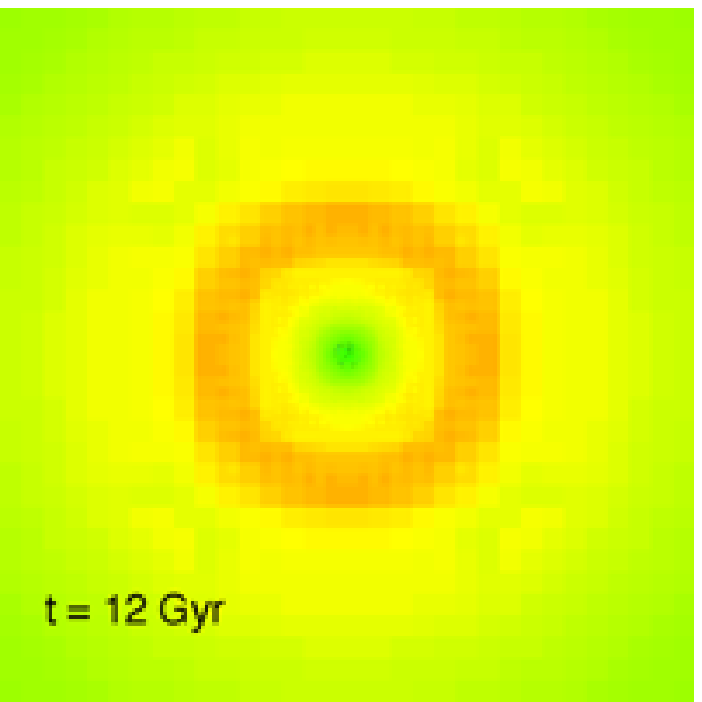,height=5.6cm,angle=0}
  }}
\end{minipage}
\begin{minipage}{5.6cm}
  \centerline{\hbox{
      \psfig{figure=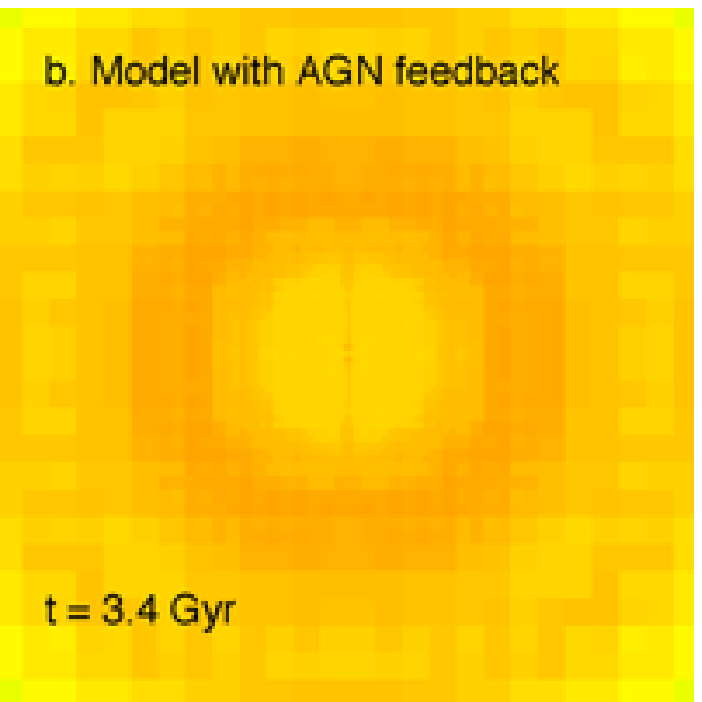,height=5.6cm,angle=0}
  }}
\end{minipage}\    \
\begin{minipage}{5.6cm}
  \centerline{\hbox{
      \psfig{figure=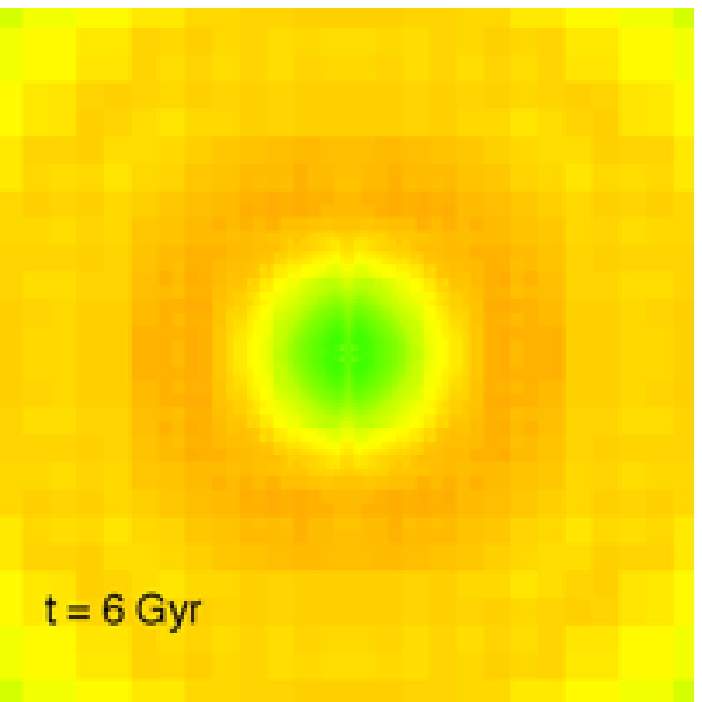,height=5.6cm,angle=0}
  }}
\end{minipage}\    \
\begin{minipage}{5.6cm}
  \centerline{\hbox{
      \psfig{figure=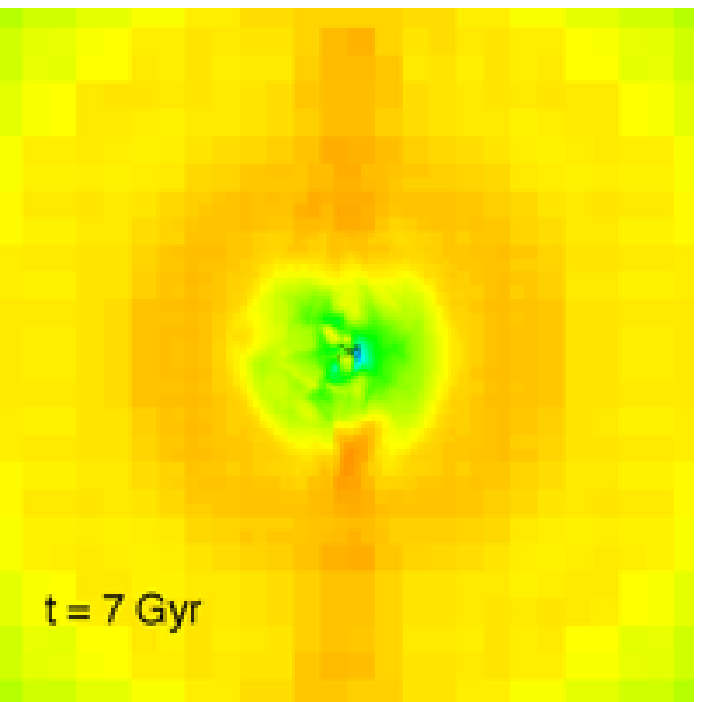,height=5.6cm,angle=0}
  }}
\end{minipage}\ \
\begin{minipage}{5.6cm}
  \centerline{\hbox{
      \psfig{figure=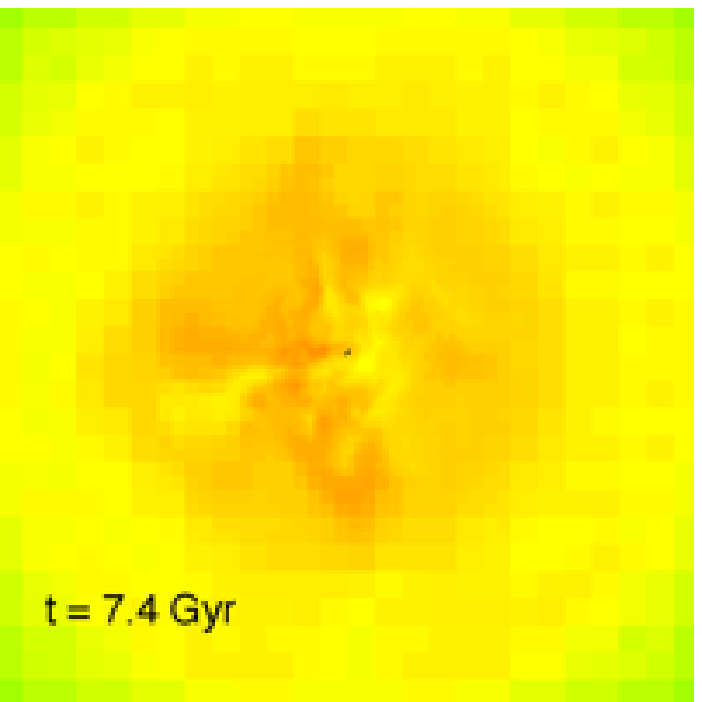,height=5.6cm,angle=0}
  }}
\end{minipage}\    \
\begin{minipage}{5.6cm}
  \centerline{\hbox{
      \psfig{figure=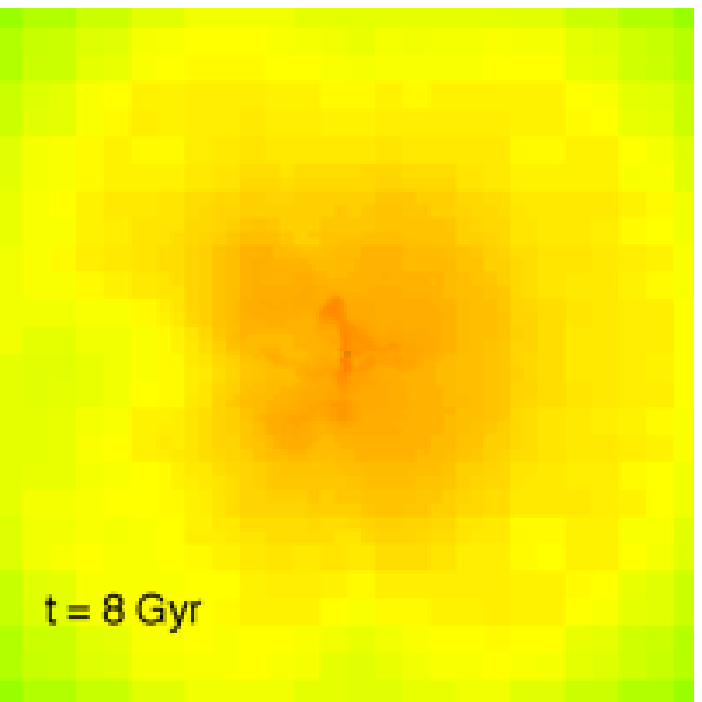,height=5.6cm,angle=0}
  }}
\end{minipage}\    \
\begin{minipage}{5.6cm}
  \centerline{\hbox{
      \psfig{figure=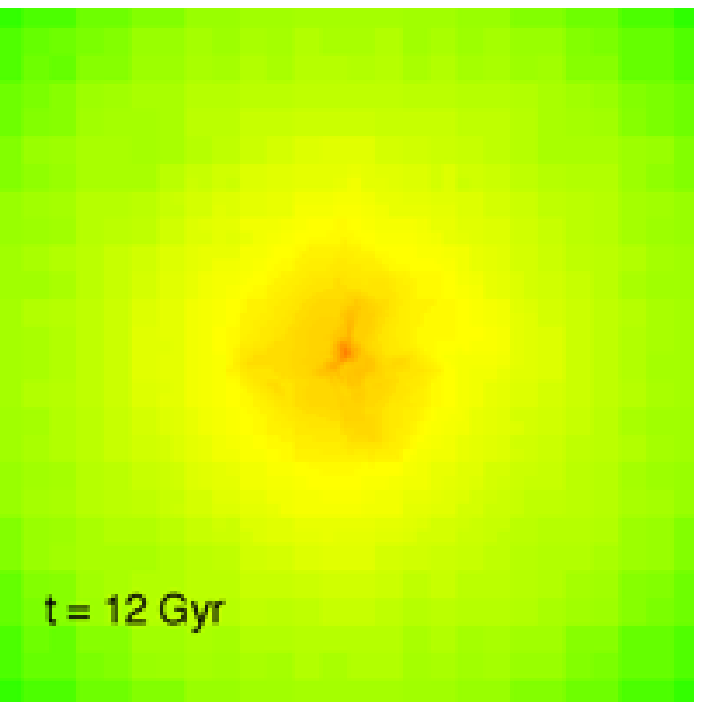,height=5.6cm,angle=0}
  }}
\end{minipage}
\caption{Maps of the emission-weighted temperature along the line of sight in the cooling flow simulation
(first row) and in the feedback simulation (second row and third row). The initial condition ($t=0$) is the same for both simulations.
Each map corresponds to a square field with a side of 260$\,$kpc. The axis of the jets is perpendicular to the line of
sight and coincides with the vertical direction on the maps. The temperature scale used to colour the snapshots is logarithmic.}
\end{figure*}

The central region that we use to compute the accretion rate of the black hole is a small cylinder aligned with the jet axis. Its radius and half-height are $r_{\rm j}=3.2\,$kpc and $h_{\rm j}=2.5\,$kpc, corresponding to $5\Delta r$ and $4\Delta r$, respectively.
The same cylinder is used to define the spatial distribution of the jet injection of mass, momentum and energy, according to
\begin{equation}
\label{supporto}
\psi={1\over 2\pi r_{\rm j}^2}{\rm exp}\left(-{x^2+y^2\over 2r_{\rm j}^2}\right){z\over h_{\rm j}^2}
\end{equation}
for $x^2+y^2\le r_{\rm j}^2$ and $|z|<h_{\rm j}$, and $\psi=0$ everywhere else.
This scheme for launching jets was introduced by \citet{omma_etal04}.
We have verified that, for the same setup, our results and theirs coincide.

\begin{figure*}
\noindent
\begin{minipage}{8.4cm}
  \centerline{\hbox{
      \psfig{figure=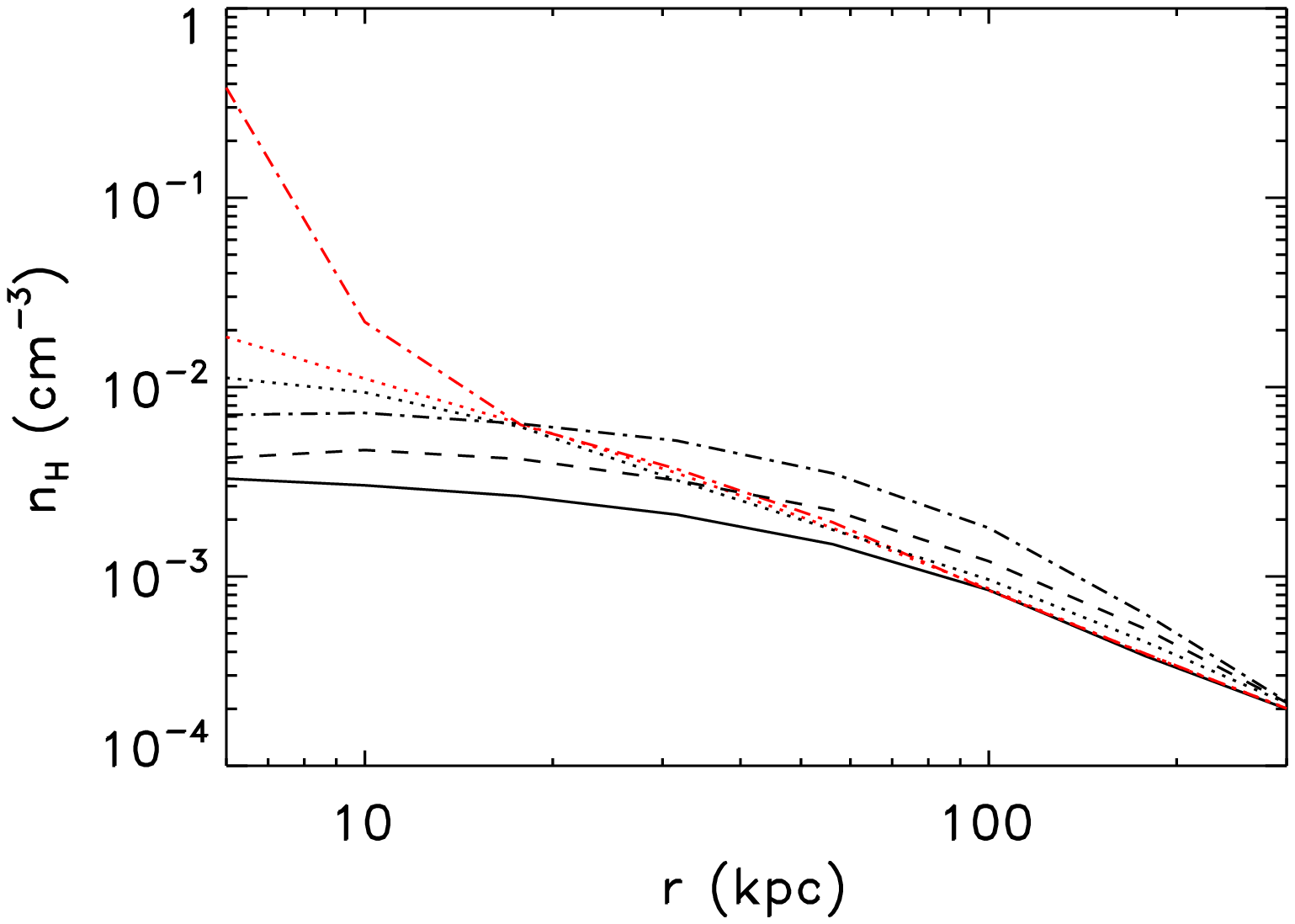,height=6.3cm,angle=0}
  }}
\end{minipage}\    \
\begin{minipage}{8.4cm}
  \centerline{\hbox{
      \psfig{figure=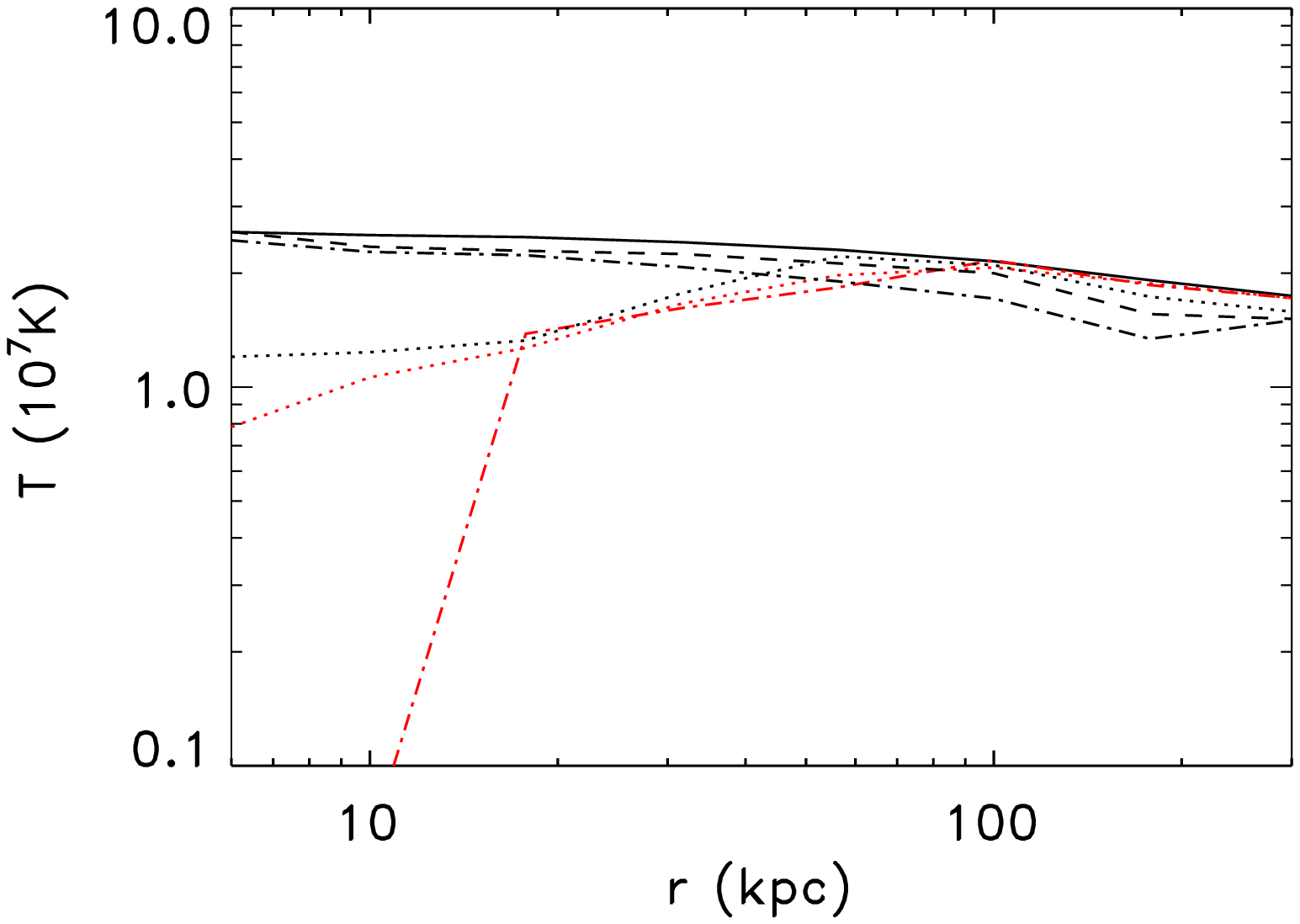,height=6.3cm,angle=0}
  }}
\end{minipage}\    \
\begin{minipage}{8.4cm}
  \centerline{\hbox{
      \psfig{figure=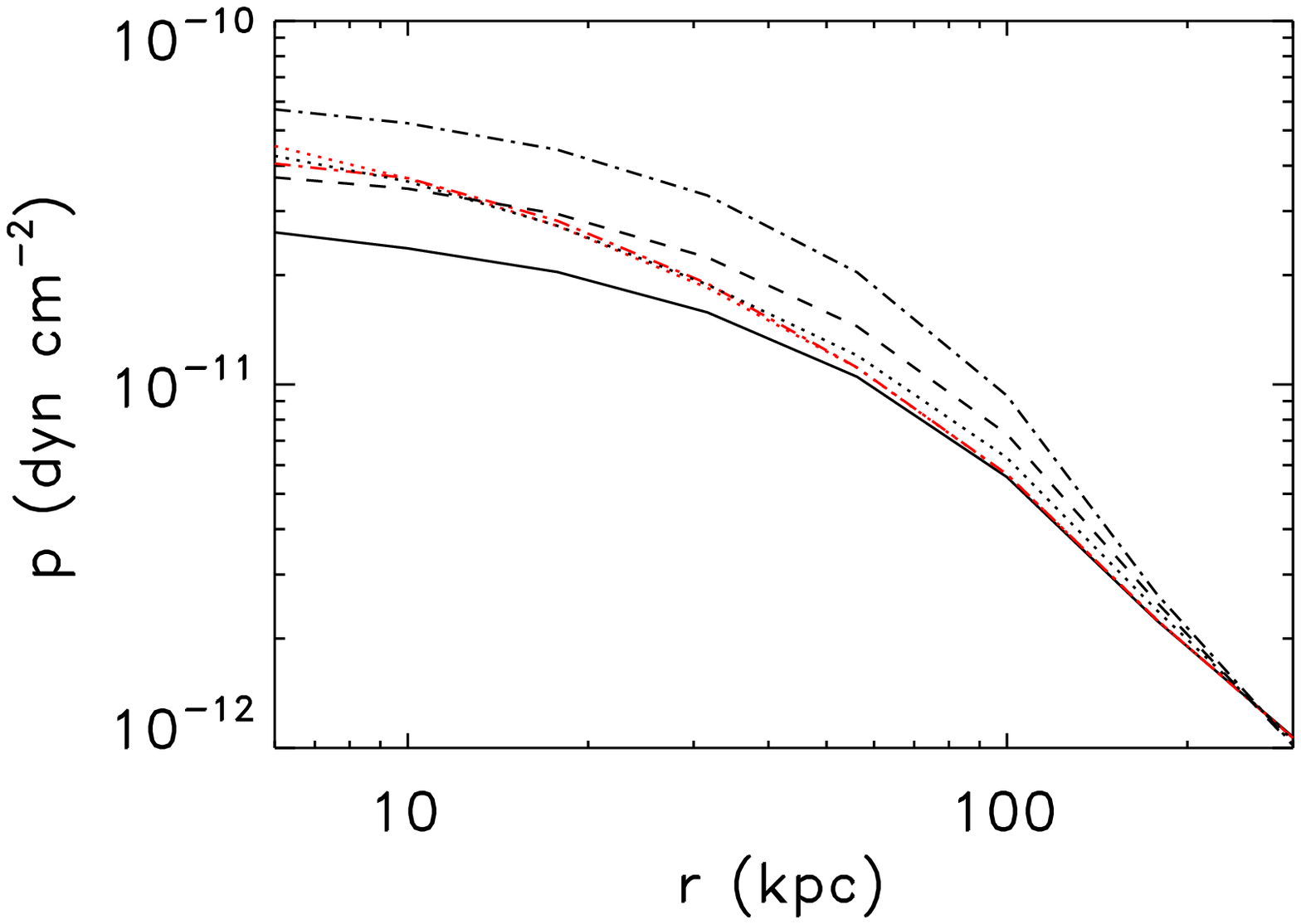,height=6.3cm,angle=0}
  }}
\end{minipage}\    \
\begin{minipage}{8.4cm}
  \centerline{\hbox{
      \psfig{figure=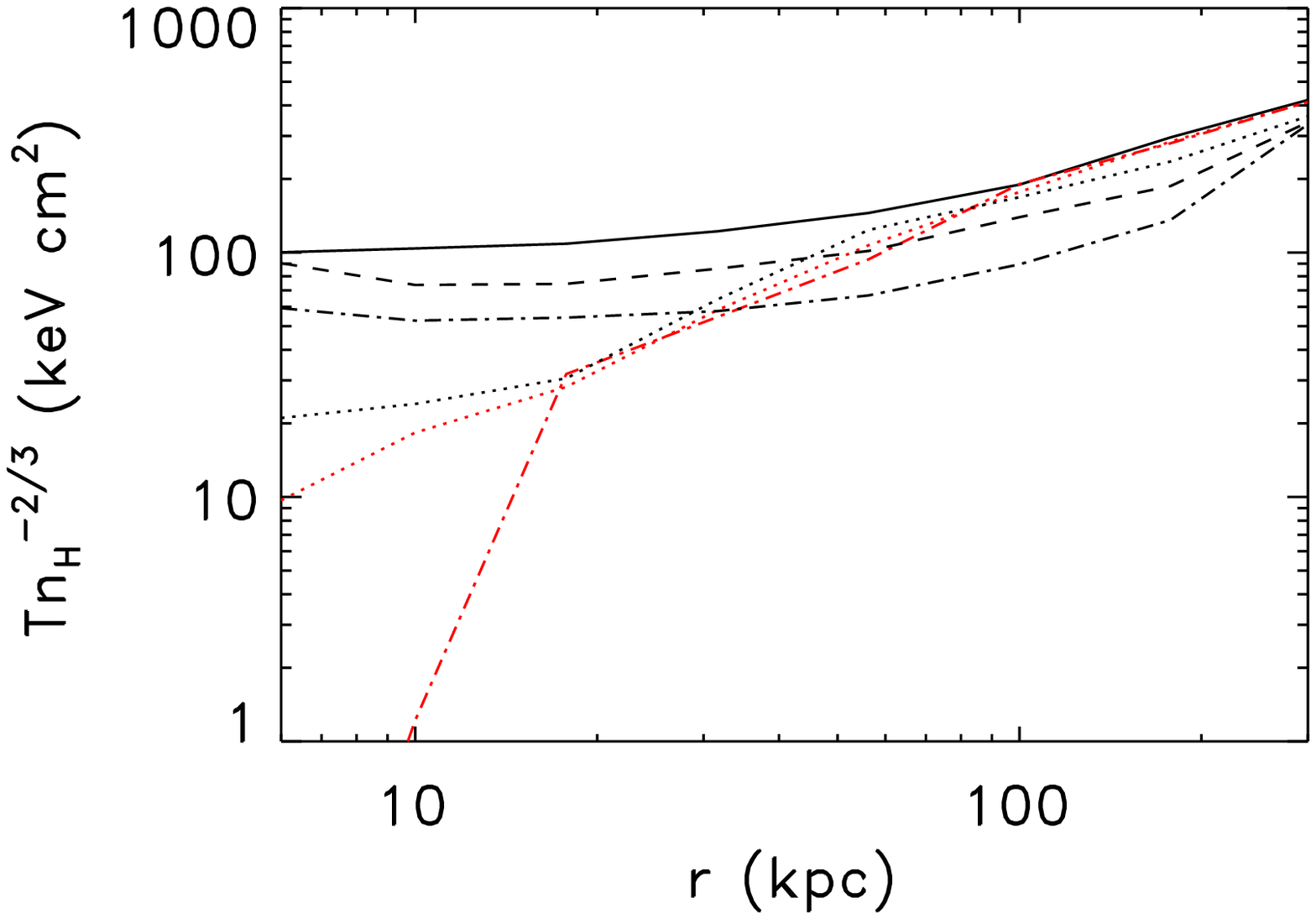,height=6.3cm,angle=0}
  }}
\end{minipage}\    \
\caption{Density, temperature, pressure and entropy profiles in the cooling flow model
(red) and in the model with AGN feedback (black) at $t\simeq 0\,$Gyr (solid lines), $t\simeq 6\,$Gyr (dotted lines), $t\simeq 8\,$ Gyr (dashed lines, only  shown for the model with AGN feedback) and $t\simeq 12\,$Gyr (dashed-dotted lines). The initial condition is the same for both the cooling flow and the AGN feedback simulation.
The density $n_{\rm H}$ is the number of hydrogen atoms divided by volume. 
$T$ is the emission-weighted temperature.
The `entropy' $Tn_{\rm H}^{1-\gamma}$ is weighted over the X-ray emission, too 
The pressure $p$ is the volume-weighted average.
All four quantities are calculated in concentrical spherical shells and are given as function of the radius
$r$ of a shell from the centre of the cluster.}
\label{profiles}
\end{figure*}

The length-scale of our cylinder is much larger than the estimate of $\sim 120\,$ pc for the Bondi radius in M87  by \citet{allen_etal06}. One could therefore reasonably worry that smoothing over such a large region may lead to an underestimate of the actual mass inflow rate at the Bondi radius. However, since the nearly flat density and temperature profiles that we measure in the central part of our simulated cluster
(see Figure~\ref{profiles}) indicate that the accretion rate computed with
Eq.~(\ref{bondi_formula}) should not be highly sensitive to the radius at which the density and temperature are evaluated.

\section{Simulations and results}

In the course of this study, we have run three simulations.
The first one is a purely hydrostatic simulation, where we have switched off cooling and feedback:
we have verified that the density distribution of the ICM remains constant for the whole duration
of the run, up to $t_{\rm sim}\sim 12\,$Gyr.
As this first simulation is just a check on the accuracy of our hydrostatic equilibrium,
we shall only comment on the other two.  The second simulation is
a standard cooling-flow simulation, where we have included cooling, but without AGN feedback.
The third simulation is our main result: both cooling and AGN feedback are considered.
The global model parameters are summarised in Table~1 and are the same for all three simulations.

In the standard cooling-flow simulation, only  at $t \sim 4\,$Gyr does the gas start to cool significantly. 
This is due to the initially high entropy of the core of the simulated cluster (e.g. \citealp{oh_benson03,mccarthy_etal04}).
However, as it starts condensing to the centre, cooling becomes catastrophic (in the bremsstrahlung regime, the cooling rate varies with the square root of the temperature, but the square of the density). At $t\sim6\,$Gyr a dense cool core is clearly visible (Fig.~1a). The temperature in the central 6$\,$kpc has
decreased by a factor of $\sim 4$ and the density has gone up by a factor of $\sim 10$ (Fig.~2).
The entropy has diminished accordingly. In Fig.~2, we have shown the entropy computed as $T n_{\rm H}^{1-\gamma}$, where $T$ is the temperature in keV and $n_{\rm H}$ is the number density of hydrogen nuclei in cm$^{-3}$. This definition is related to
the one given in Section~2.2 by $s= kT(n_{\rm H}\mu m_{\rm p})^{1-\gamma}$.
By the end of the simulation, at $t=12\,$Gyr, the central density is more than a hundred times higher than its  initial value, the central temperature and entropy have decreased dramatically,
while the pressure has grown by a factor of $\sim 2$.
The pressure has changed more moderately than the other three quantities  because 
the cooling time is longer than the dynamical time and the ICM has cooled quasi-statically.

The gas in our simulations is either hot, with a temperature of the order of the virial temperature of the halo ($T\sim 10^7\,$K), or cold ($T\sim 10^4\,$K), with a negligible fraction of the mass at intermediate values. It is therefore straightforward to separate the cold gas from the hot gas and to see that
a large mass of cold gas builds up in the central few kiloparsecs.
This mass  is of $\sim 10^{10}M_\odot$ at $t\sim6\,$Gyr.
By the end of the simulation, it grows above $4 \times 10^{11}M_\odot$ (Fig.~3).
This cannot be appreciated from Fig.~1a because gas that cools below $10^6\,$K goes down to $10^4\,$K very rapidly, fragments into clumps and it gets squeezed into a few
dozens of cells. They contain all the cold gas and occupy a central region with a
radius of $\sim 4\,$kpc. The lack of net angular momentum in our initial gas distribution,
the absence of a prescription for star formation and supernova feedback
and the limited number of cells in this central region imply  that we have evolved our cluster to the 
limit where our physical description breaks down.

A resolution study shows that the reason why only $\sim 3.5\%$ of the baryons cool is that
our simulation is not fully converged but achieving the convergence resolution is impossible with the short timesteps of the  jet simulation. We have therefore shown the outputs of the cooling flow simulation at the same resolution used for the feedback simulation. We shall discuss later how this is likely to affect our conclusions.

In the simulation with feedback, the interaction of the black hole with the ICM starts gently because
the initial accretion rate, $\dot{M}_\bullet \sim 5\times 10^{-4} M_\odot{\rm\,yr}^{-1}$, is low (Fig.~4).
This is due to the high entropy of the cluster core in the initial condition.
The AGN empties a narrow tube along the jet axis (Fig.~1b and Fig.~5).
The jets look like spring onions, with a thin round part at the bottom, which continues into a stem.
This can be explained by the pressure of the jets when they come out 
of the injection zone, which is slightly lower than the pressure in the cluster core. So the jets are squeezed at the sides, until they reach the pressure of the ICM.
Due to the continuity equation, the shrinking of the pipe's section is accompanied by an increase of
the flow speed. This is seen as an increase of the ram pressure $\rho v^2$ at the points where the jets
on the temperature maps become narrower (Fig.~5).

Using directly the density and temperature maps, we can determine that the jets are propagating into the cluster core at a subsonic speed of about $500{\rm\,km\,s}^{-1}$. This subsonic propagation can also be demonstrated through the absence of hotspots in the lobes. The temperature of the plasma in the jets has a maximum at the central source and decreases moving outwards. This is consistent with the notion that we are simulating  a Fanaroff-Riley type I or core-dominated radio source. The behaviour of the density mirrors that of the temperature. The density grows outwards along the jets, as the outflowing plasma slows down and gathers in the jet lobes, where it mixes with entrained material.

At early times, the AGN has little effect on the state and evolution of the ICM in the cluster core. The gas entropy in the cluster core is rather high, resulting in a rather low black hole accretion rate of
$\dot{M}_\bullet \lsim10^{-3}M_\odot{\rm\,yr}^{-1}$.  Therefore, even with the high efficiency we have considered in the accretion-ejection conversion process,  the power injected by the AGN into the ICM is less than the rate at which the ICM looses energy through radiative cooling (Fig.~4). Still, the injected energy delays the onset of the fast cooling phase from $t\sim 4\,$Gyr up to $t\sim 6\,$Gyr (see Figs.~3). Another reason for this low overall efficiency at early time is that the  high entropy plasma in the jets flows outwards, rather than enriching the entropy of the ICM around the black hole. This can be seen by comparing the cooling flow simulation and the AGN feedback simulation using Figure~1at $t=6\,$Gyr. Both contain a cool core at $T\lsim 10^7\,$K, but the most visible difference is  at $r\gsim 100\,$kpc, where the ICM has been heated more efficiently by the AGN feedback than in the pure cooling flow simulation.

By $t\simeq 6.5\,$Gyr, the cooling catastrophe, although it was delayed by almost 2 Gyr, still occurs.  As the gas in the cluster core is cooling and condensing faster and faster, the accretion rate of the black hole grows by two orders of magnitudes in $\sim 0.5\,$Gyr, and reaches a peak of $\dot{M}_\bullet \sim 0.1M_\odot{\rm\,yr}^{-1}$ at $t\simeq 7\,$Gyr. At this rate, the jet power, $P_{\rm j}^{\rm max}\sim 10^{45}{\rm\,erg\,s}^{-1}$, exceeds the bremsstrahlung luminosity of the cluster by a factor of $\sim 10^2$. The thermal pressure $p$ at the base of the jets and the ram pressure $\rho v^2$ throughout the length of the jets are no longer negligible compared to the pressure of the ICM in which the jets propagate. 

Given the new physical state in the cluster core, the jet propagation is no longer stable. The Kelvin-Helmoltz instability sets in and rapidly amplify small scale fluctuations already present in the flow. This effect is so strong that the flow at $t=6\,$Gyr is no longer axisymmetric, although our initial conditions were perfectly axisymmetric (Fig.~5).
A complete analytical description of jet instabilities is beyond the scope of this paper (see \cite{smith_etal83} for a comprehensive study). 
Our numerical experiment suggests that 
these instabilities grow faster in the phase of rapid accretion: 
the jets become turbulent and break up into bubbles, which, because of their lower density than the ICM, are buoyantly driven outside the cluster core. In this turbulent atmosphere, bubbles may rise in directions not aligned with the jets' axis. Fig.~6 shows such an extreme situation: the jet pointing downwards blows a bubble that rises perpendicular to the jet.
The hydrodynamic interaction of the jets with the ICM is, therefore, a complicated process with  many different aspects to it. The jets inflate cavities and through their expansion blow the gas in the centre of the cluster away. They heat gas through shocks, but also through turbulent mixing of jet material with the ICM.  AGN feedback reheats most of the cold gas that has accumulated at the centre of the cluster during the cooling catastrophe and causes $\dot{M}_\bullet$ to drop by a factor of $\sim 50$ at $7{\rm\,Gyr}<t<8{\rm\,Gyr}$. After reaching its maximum value of $\dot{M}_\bullet \sim2 \times 10^{-1}M_\odot{\rm\,yr}^{-1}$, the accretion curve is reasonably modelled by a decreasing exponential, with an e-folding time scale of $t_{\rm life}\sim 2.5\times 10^8\,$yr, consistent with the typical life time of a radio source.

\begin{figure}
\centerline{\hbox{
\psfig{figure=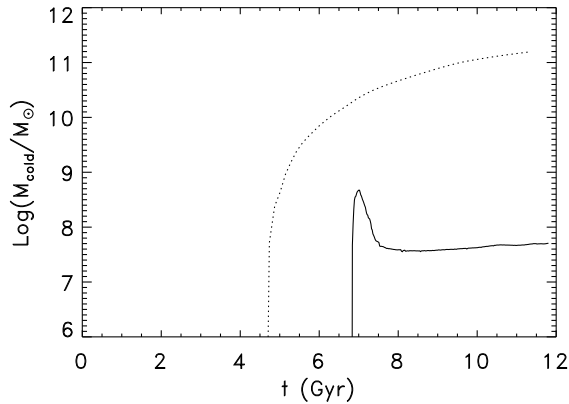,height=6cm,angle=0}}}
\caption{The mass of the cold gas ($T<10^6\,$K) in the computational volume as a function of time.
The dotted line corresponds to the cooling flow simulation and the solid line to the feedback simulation.}
\end{figure}

\begin{figure}
\centerline{\hbox{
\psfig{figure=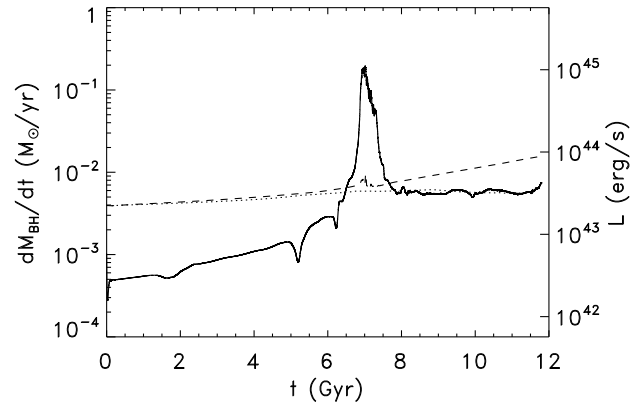,height=6cm,angle=0}}}
\caption{The accretion rate of the black hole, $\dot{M}_\bullet$, and the corresponding jet mechanical luminosity,
$0.1\dot{M}_\bullet{\rm c}^2$ in the simulation with AGN feedback (solid line). The dotted line and the dashed line show the X-ray luminosity emitted by the ICM in the 
cooling flow simulation and the feedback simulation, respectively.}
\end{figure}

At $t\gsim 8{\rm\,Gyr}$, the central temperature of the hot gas, the mass of the cold gas and the black hole's accretion rate settle respectively to $T\sim 3\times 10^7\,$K, $M_{\rm cold}\sim 5\times 10^7M_\odot$ and $\dot{M}_\bullet\sim 2\times 10^{-3}$, without substantial variations until the end of the simulation at $t\simeq 12\,$Gyr (Figs.~2,~3 and~4, respectively). This strongly suggests that an equilibrium state has been reached. 
This new equilibrium regime can be characterised by a highly turbulent and highly pressurised state within the cluster core. These new properties play an important role in the system's self-regulation.
Before the cooling flow goes unstable, the hot plasma flowing out of the central source easily escapes the cluster core, funnelled through the high density ICM by a straight and empty channel excavated by Kelvin-Helmholtz stable jets. In the new equilibrium state, this is no longer the case: the jets must work constantly to open a path for themselves. 
\citet{omma_binney04} had also pointed out the importance of closing the channel for effective feedback.
Moreover, the higher pressure of the ICM contributes to confine the outflow within the central region and force most of the mechanical energy to remain inside the core. Therefore, the jets at $t=12\,$Gyr are slower, hotter and ultimately more effective than those at $t=6\,$Gyr (Fig.~5): a higher fraction of the AGN power is converted into heat that compensate bremsstrahlung cooling in the central region. This explain why the inner region of the cluster has reached this new hydrostatic and quasi-adiabatic regime.

In the catastrophic cooling phase ($t \sim 6\,$Gyr), the entropy of the ICM within the core decreases from $100\,{\rm keV}{\rm\,cm}^2$ to $20\,{\rm keV}{\rm\,cm}^2$. At $t\sim 8\,$Gyr, after the strong AGN outburst, the central entropy is nearly back to its initial value. By $t\sim 12\,$Gyr, the entropy of the cluster core has started to decrease again, albeit by a small amount and over a rather long period of several Gyr. This slow evolution is due to a secular rise of the central density, as the central temperature remains quasi constant (Fig.~2). 
In Section~4 we shall argue that the temperature does not change much
because it adjusts itself to the value at which AGN heating balances radiative cooling in the cluster core.
We should make note of the fact that the profiles in Fig.~2 are useful to understand the physical processes in the simulated cluster but cannot be taken as accurate predictions to be compared with observations because of the idealised gravitational potential used in the simulations,
which neglects the dominant stellar component at $\lsim 10\,$kpc (e.g. \citealp{mamon_lokas05}). We have also neglected
the role of cosmic rays as an additional non-equilibrium component, whose observational signature mainly resides in the radio emission of galaxy clusters. 

\begin{figure*}
\noindent
\begin{minipage}{5.6cm}
  \centerline{\hbox{
      \psfig{figure=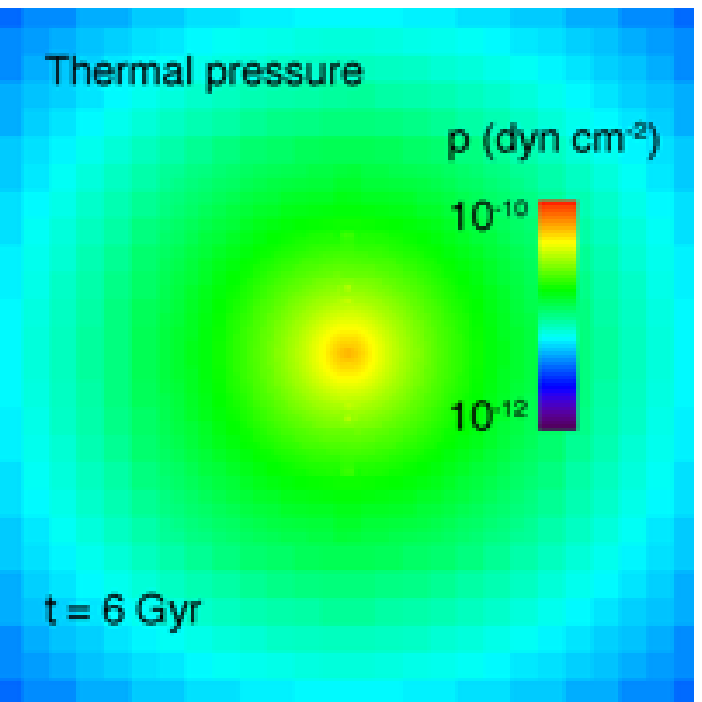,height=5.6cm,angle=0}
  }}
\end{minipage}\    \
\begin{minipage}{5.6cm}
  \centerline{\hbox{
      \psfig{figure=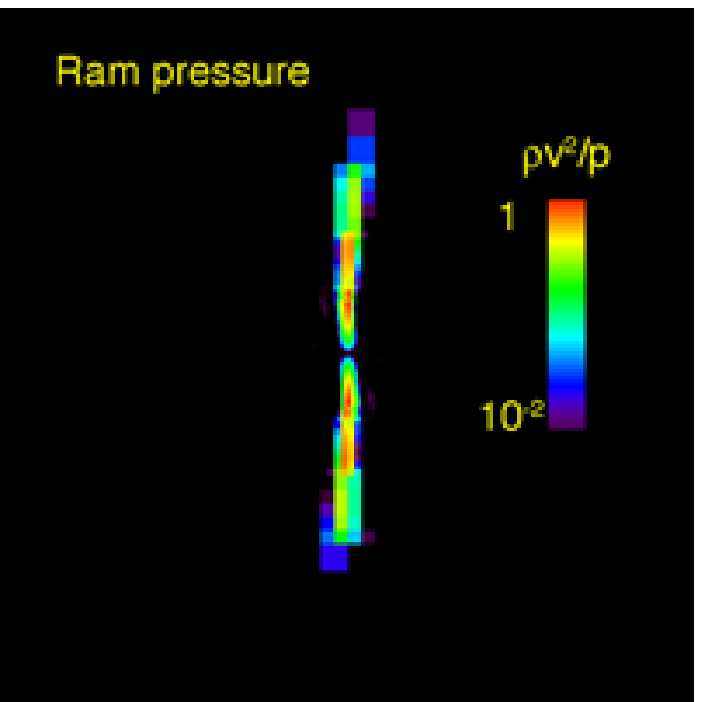,height=5.6cm,angle=0}
  }}
\end{minipage}\    \
\begin{minipage}{5.6cm}
  \centerline{\hbox{
      \psfig{figure=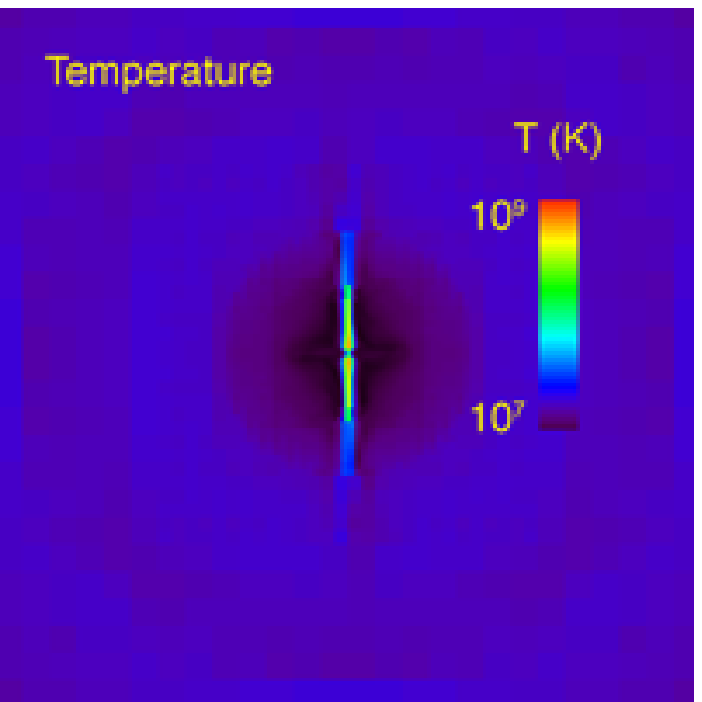,height=5.6cm,angle=0}
  }}
\end{minipage}
\begin{minipage}{5.6cm}
  \centerline{\hbox{
      \psfig{figure=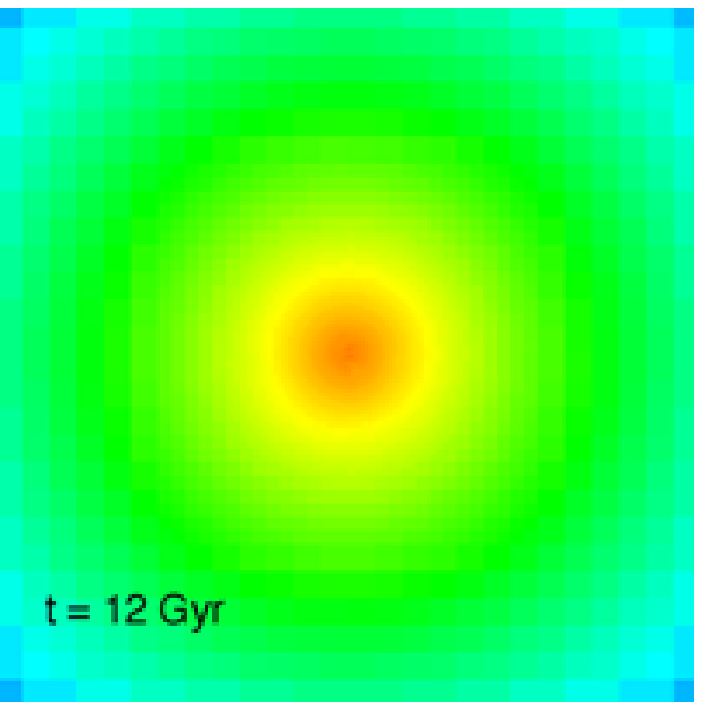,height=5.6cm,angle=0}
  }}
\end{minipage}\    \
\begin{minipage}{5.6cm}
  \centerline{\hbox{
      \psfig{figure=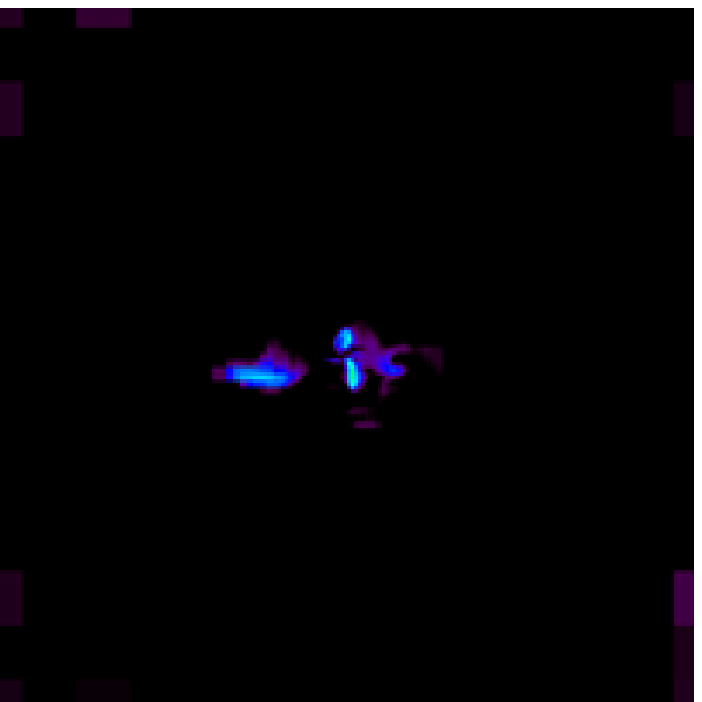,height=5.6cm,angle=0}
  }}
\end{minipage}\    \
\begin{minipage}{5.6cm}
  \centerline{\hbox{
      \psfig{figure=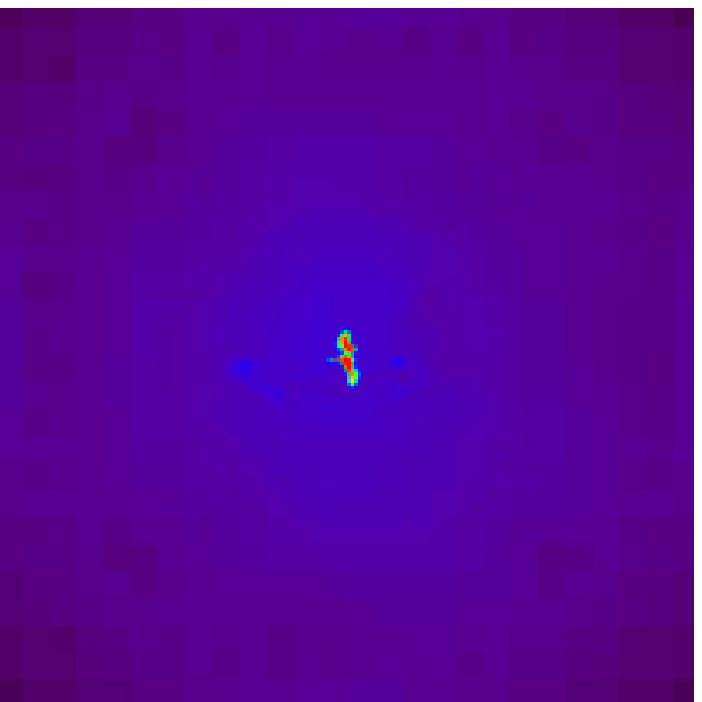,height=5.6cm,angle=0}
  }}
\end{minipage}
\caption{The thermal pressure, $p=\rho kT/\mu m_{\rm p}$,  the ratio of the ram pressure to the thermal pressure, 
$\rho v^2/p$, and the temperature, $T$, of the gas at two different times ($t=6\,$Gyr in the upper row and $t=12\,$Gyr in the lower one)  on a section that contains the jet axis and is orthogonal to the line of sight used to plot Fig.~1.
The charts have a side of $260\,$kpc. All colour scales are logarithmic.}
\end{figure*}

The cooling of gas in the outer parts of the cluster core causes the cluster gas to flow inwards and slowly compress the central region (Fig.~1). This secular compression is controlled by the quasi-adiabatic equilibrium we have reached within the cluster core. In a more realistic cosmological environment, the merging of clumpy satellites of cold gas would have triggered another phase of strong AGN activity, followed by another long period of quasi-equilibrium.

\section{Discussion and conclusion}
 
Our views on the thermal evolution of the hot gas in galaxy clusters have changed considerably in recent years due to data from CHANDRA and XMM \citep{fabian_etal03}.
These X-ray satellites have found no evidence for gas below one third of the virial temperature
\citep{allen_etal01,peterson_etal03}.
Millimetre observations of CO lines with the IRAM 30 Metre Telescope, the James 
Clerk Maxwell Telescope and the Caltech Submillimeter Observatory have detected the elusive cold molecular gas  (\citealp{salome_etal06} and references therein). However,
measured cooling rates are substantially lower than inferred from  X-ray luminosity
 (also see \citealp{bregman_etal06}). Inefficient cooling in massive haloes is also necessary to explain
 several galaxy properties \citep{cattaneo_etal06,cattaneo_etal07,croton_etal06,bower_etal06}.

There is increasing consensus that AGN feedback is  the reason
why the cooling rate is much lower than one would expect (e.g.
\citealp{tabor_binney93,binney_tabor95,tucker_david97,ciotti_ostriker97,ciotti_ostriker01,churazov_etal01,quilis_etal01,reynolds_etal01,reynolds_etal02,basson_alexander03,begelman04,omma_etal04,omma_binney04,ruszkowski_etal04,zanni_etal05,brighenti_mathews06,fabian_etal06,mathews_etal06,nusser_etal06,sijacki_springel06,vernaleo_reynolds06}).
Non-gravitational heating is also needed to explain the high entropy floor of the ICM and why the observed relation between X-ray temperature and X-ray luminosity deviates from the virial relation
(e.g. \citealp{mccarthy_etal02,roy_etal04,borgani_etal05}).
AGN feedback can be radiative and mechanic. Most of the computer simulations that
follow the hydrodynamics of the interaction of AGNs with the ICM  have concentrated on mechanic feedback because of the
direct observational evidence that jets from AGNs can open large cavities in the ICM
\citep{mcnamara_etal05,fabian_etal06}.

However, while substantial work has been put into understanding the mechanisms through which jets affect the ICM, the exploration of the long-term outcome of this interaction and of the feedback that it returns to the central engine has just started.
\citet{omma_binney04} compared two simulations of a cooling flow cluster in which  jets were turned on at different times. They assumed that a later activation of the central source results in more powerful jets, since the central density of the ICM is higher at a more advanced stage of the cooling
catastrophe. They used this as an argument to
suggest that the interaction of the black hole with the ICM is structurally stable. 
However, in nearly all the studies done until now, the power and the duration of the jets were set by hand following observational estimates. 
\citet{brighenti_mathews06}'s 2D simulations and \citet{vernaleo_reynolds06}'s 3D study,
whose simulations did not last long enough to attain the self-regulated regime described in this article,
 are the only ones so far who have attempted to relate the properties of jets to the cooling rate of the ICM.

This paper fills this gap in our comprehension of the problem by closing the feedback loop  with a 
self-consistent model of the interaction between the black hole and the ICM.
The accretion rate of the black hole is computed from the density and temperature of the ICM.
The power of the jets is proportional to the accretion rate of the black hole and the jets affect back the accretion rate of the black hole by changing the properties of the ICM.

The specific model used to calculate the accretion rate of the black hole is the one proposed by \citet{bondi52}.
In this model, which corresponds to spherical accretion of gas that is supported by thermal pressure and
static at infinity, the accretion rate of the black hole is $\dot{M_\bullet}\propto M_\bullet^2s^{-3/2}$
where $s\propto p\rho^{-5/3}\propto T n_{\rm H}^{-2/3}$ is a measure of the entropy of the gas around the black hole (Section~2.2). 

 In this work, we have simulated the evolution of the ICM in a halo of $M_{\rm halo}=1.5\times 10^{14}M_\odot$ described by the NFW density profile \citep{navarro_etal97} and containing a central black hole of $M_\bullet=3\times 10^9M_\odot$ starting from hydrostatic initial conditions.
In our simulations,
the growth of the black hole is so small that $M_\bullet$ is nearly constant, but its value is important because it determines the power of the central source. 
The black hole feeds energy back to the ICM 
by the injection of hot plasma with an outflow speed of $v_{\rm j}\sim 1400{\rm\,km\,s}^{-1}$)
in two symmetric volumes at a small distance from the black hole.  

From our simulations of the interaction of the black hole with the ICM, we learn  two new fundamental results. In presence of a massive black hole, a cooling-flow cluster switches on an  AGN of $\sim 10^{45}{\rm\,erg\,s}^{-1}$soon after the onset of the cooling catastrophe. Its lifetime is short,  $\sim 2.5\times 10^8\,$yr, because twin jets blow away and reheat the gas in the cluster core. After the AGN phase, there are at least $4\,$Gyr during which the accretion rate of the black hole and the mass of the cold gas in the cluster core stay constant, suggesting
that a self-regulated equilibrium has been reached.

Turbulence stirred by jet instabilities mixes the jet plasma with the ICM in the cluster core,
as previously found by \citet{churazov_etal01}, \citet{quilis_etal01} and \citet{reynolds_etal02}.
A highly turbulent and highly pressurised state of the ICM is essential for the effectiveness of our feedback mechanism.
Without it, the jet energy would have left the cluster core and the AGN would have not heated the cooling gas efficiently. 
Before the jets grew unstable, most of the power was indeed deposited far from the centre. 
Self-regulation was then impossible.

\begin{figure}
\noindent
\begin{minipage}{8.4cm}
  \centerline{\hbox{
      \psfig{figure=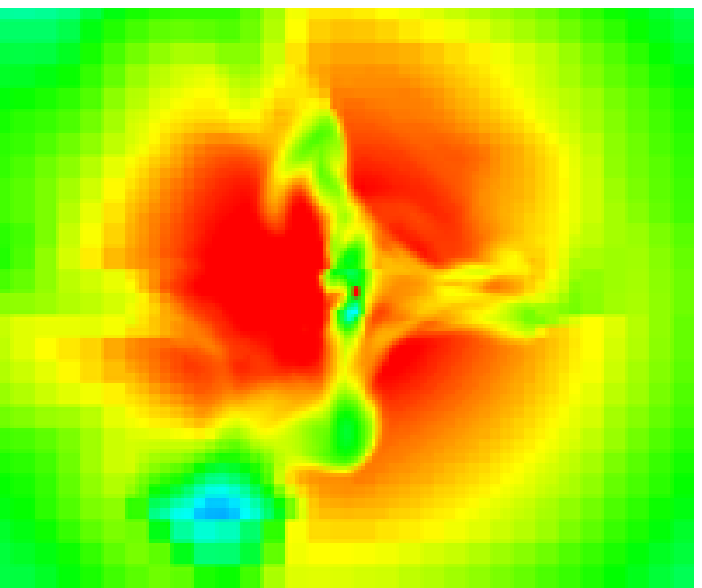,width=7.3cm,angle=0}
  }}
\end{minipage}
\begin{minipage}{8.4cm}
  \centerline{\hbox{
      \psfig{figure=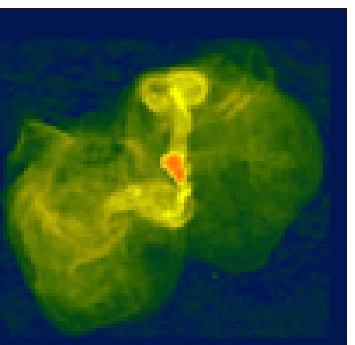,width=7.3cm,angle=0}
  }}
\end{minipage}\    \
\caption{The cavities that AGN feedback has opened in the hot gas appear as lower density regions.
Above is a density map extracted from the simulation at $t\sim 8\,$Gyr.
Below is an observational image of the radio emission from the synchrotron plasma that fills the X-ray cavities in M87 \citep{owen_etal00}.
The two figures have the same physical size ($\sim 70{\rm\,kpc}\times 70{\rm\,kpc}$) and show an interesting morphological similarity.}
\end{figure}

 The violent outburst following the cooling catastrophe causes the thermal and turbulent pressure to rise significantly, until most of the jet energy is confined inside the cluster core. This confinement allows energy released by the black hole to have an immediate feedback on its accretion rate.
This ensures that an equilibrium is reached between the accretion rate of the black hole and the temperature of the surrounding gas. 

The equilibrium state can be determined by balancing AGN heating with radiative cooling in the cluster core, assuming that it is now a closed system, so that all the energy released by the AGN remains within the boundaries of the core. Eqs.~(5) and~(8) give the heating rate $P_{\rm j} \simeq 4\pi\epsilon{\rm c}^2({\rm G}M_\bullet)^2m_{\rm p}n_{\rm H}c_{\rm s}^{-3} \propto n_{\rm H} T^{-3/2}$, while the total radiated power is given by $L_{\rm  X}=n_{\rm H}\Lambda(T)M_{\rm H}/m_{\rm p} \propto n_{\rm H} T^{1/2}$, where $M_{\rm H}$, the total hydrogen mass in the core, is assumed constant ($M_{\rm H}\sim 10^{11}M_\odot$). Given $M_{\rm H}$ and $M_\bullet$, there is one equilibrium temperature for which $P_{\rm j}=L_{\rm X}$ and its value its independent of $n_{\rm H}$. From this simple argument, we compute $T = 2 - 3\times10^7\,$K. This figure agrees with the results of our simulations (Fig.~2) and is slightly higher than the virial temperature of our simulated cluster,
$T_{\rm vir}\simeq 1.7\times 10^7\,$K.

If the mass of the black hole had been significantly lower than the value considered here, the equilibrium temperature would have been much lower than the virial temperature. As,
by definition, the virial temperature is that at which thermal pressure balances the gravitational force,  
AGN heating would not have been able to stop the contraction of the cluster core with such black hole mass.
The black hole would have had to grow considerably for feedback eventually to become efficient.

In addition, if the local cooling rate exceeds the local heating rate around the black hole, the net effect is cooling, the temperature decreases, the density increases and the heating rate goes up. If, on the other hand, the heating rate becomes larger than the cooling rate, than the temperature increases, the density decreases and the heating rate goes down. This thermal equilibrium is, therefore, likely to remain stable and self-regulated for several Gyr, as our simulations have demonstrated. 
The infall of cold gas clumps from outside the central region
(such as those observed by \citealp{salome_etal06})
 may eventually give rise to a new period of AGN activity.

The simulations presented in this paper prove that the interaction of the AGN with the cooling flow evolves naturally towards a self-regulated equilibrium state after a short phase of strong activity. 
We suspect that this feature is robust and does not strongly depends on the mechanism that thermalizes
the power channelled into the jets.

Our simulations do not provide an accurate description of an individual object such as M87, since, among other approximations, we have neglected the stellar contribution, which dominates the gravity in the central few kiloparsecs. However, the halo and the black hole parameters were chosen to mimic the values appropriate for M87. Therefore, a careful comparison with M87 is useful because it gives us an idea to what extent this simple model represents the physical reality. 

\citet{boehringer_etal94} estimate from ROSAT data that that the total luminosity of the Virgo cluster in the $0.1-2.4\,$keV band is $L_X\sim 8\times 10^{43}{\rm erg\,s}^{-1}$. Only $70\%$ of this power comes from the hot gas in the halo of M87, while most of the rest comes from M49 and M86. Note that this observational estimate is compensated by the fact that another $50\%$ of $L_X$ comes from outside the observation band.  The final luminosity of our simulated cluster, $L_X\sim 9\times 10^{43}{\rm erg\,s}^{-1}$, is therefore very close to value inferred from observations of M87. To our opinion, it is even more significant that the black hole accretion rate we have computed is consistent with the value inferred by \citet{dimatteo_etal03} from studying M87 with the CHANDRA X-ray images: they found $\dot{M}_\bullet\sim 0.1M_\odot{\rm\,yr}^{-1}$ and $L_{\rm Bondi}\sim 5\times 10^{44}{\rm\,erg\,s}^{-1}$.There are also significant morphological similarities between the cavities opened by the jets in our simulation and the radio structures observed in M87 by \citep{owen_etal00} (see Figure 6). The duration of the active phase, $t_{\rm life}\sim 2.5\times 10^8\,$yr, is also in the range of what is expected for the typical lifetime of a radio source.

\section{Acknowledgements}

The authors thank J. Binney, F. Combes, A. Dekel, S. Faber, W. Forman, M. Hoeft, A. Klypin  and J. Roland for interesting discussion. The simulations were performed at the CNRS Supercomputing Centre IDRIS and at the CEA Supercomputing Centre CCRT. This work has been supported by the Horizon Project.

\bibliographystyle{mn2e}
\bibliography{references}

\begin{thebibliography}{}

\bibitem[\protect\citeauthoryear{{Allen}, {Dunn}, {Fabian}, {Taylor} \&
  {Reynolds}}{{Allen} et~al.}{2006}]{allen_etal06}
{Allen} S.~W.,  {Dunn} R.~J.~H.,  {Fabian} A.~C.,  {Taylor} G.~B.,
  {Reynolds} C.~S.,  2006, \mnras, 372, 21

\bibitem[\protect\citeauthoryear{{Allen}, {Fabian}, {Johnstone}, {Arnaud} \&
  {Nulsen}}{{Allen} et~al.}{2001}]{allen_etal01}
{Allen} S.~W.,  {Fabian} A.~C.,  {Johnstone} R.~M.,  {Arnaud} K.~A.,
  {Nulsen} P.~E.~J.,  2001, \mnras, 322, 589

\bibitem[\protect\citeauthoryear{{Arnaud}, {Fabian}, {Eales}, {Jones} \&
  {Forman}}{{Arnaud} et~al.}{1984}]{arnaud_etal84}
{Arnaud} K.~A.,  {Fabian} A.~C.,  {Eales} S.~A.,  {Jones} C.,    {Forman} W.,
  1984, \mnras, 211, 981

\bibitem[\protect\citeauthoryear{{Babul}, {Balogh}, {Lewis} \& {Poole}}{{Babul}
  et~al.}{2002}]{babul_etal02}
{Babul} A.,  {Balogh} M.~L.,  {Lewis} G.~F.,    {Poole} G.~B.,  2002, \mnras,
  330, 329

\bibitem[\protect\citeauthoryear{{Barger}, {Cowie}, {Mushotzky}, {Yang},
  {Wang}, {Steffen} \& {Capak}}{{Barger} et~al.}{2005}]{barger_etal05}
{Barger} A.~J.,  {Cowie} L.~L.,  {Mushotzky} R.~F.,  {Yang} Y.,  {Wang} W.-H.,
  {Steffen} A.~T.,    {Capak} P.,  2005, \aj, 129, 578

\bibitem[\protect\citeauthoryear{{Basson} \& {Alexander}}{{Basson} \&
  {Alexander}}{2003}]{basson_alexander03}
{Basson} J.~F.,  {Alexander} P.,  2003, \mnras, 339, 353

\bibitem[\protect\citeauthoryear{{Begelman}}{{Begelman}}{2004}]{begelman04}
{Begelman} M.~C.,  2004, in {Ho} L.~C.,  ed., Coevolution of Black Holes and
  Galaxies {AGN Feedback Mechanisms}.
p.~374

\bibitem[\protect\citeauthoryear{{Binney} \& {Tabor}}{{Binney} \&
  {Tabor}}{1995}]{binney_tabor95}
{Binney} J.,  {Tabor} G.,  1995, \mnras, 276, 663

\bibitem[\protect\citeauthoryear{{Blandford} \& {Begelman}}{{Blandford} \&
  {Begelman}}{1999}]{blandford_begelman99}
{Blandford} R.~D.,  {Begelman} M.~C.,  1999, \mnras, 303, L1

\bibitem[\protect\citeauthoryear{{Bohringer}, {Briel}, {Schwarz}, {Voges},
  {Hartner} \& {Trumper}}{{Bohringer} et~al.}{1994}]{boehringer_etal94}
{Bohringer} H.,  {Briel} U.~G.,  {Schwarz} R.~A.,  {Voges} W.,  {Hartner} G.,
   {Trumper} J.,  1994, \nat, 368, 828

\bibitem[\protect\citeauthoryear{{Bondi}}{{Bondi}}{1952}]{bondi52}
{Bondi} H.,  1952, \mnras, 112, 195

\bibitem[\protect\citeauthoryear{{Borgani}, {Finoguenov}, {Kay}, {Ponman},
  {Springel}, {Tozzi} \& {Voit}}{{Borgani} et~al.}{2005}]{borgani_etal05}
{Borgani} S.,  {Finoguenov} A.,  {Kay} S.~T.,  {Ponman} T.~J.,  {Springel} V.,
  {Tozzi} P.,    {Voit} G.~M.,  2005, \mnras, 361, 233

\bibitem[\protect\citeauthoryear{{Bower}, {Benson}, {Malbon}, {Helly}, {Frenk},
  {Baugh}, {Cole} \& {Lacey}}{{Bower} et~al.}{2006}]{bower_etal06}
{Bower} R.~G.,  {Benson} A.~J.,  {Malbon} R.,  {Helly} J.~C.,  {Frenk} C.~S.,
  {Baugh} C.~M.,  {Cole} S.,    {Lacey} C.~G.,  2006, \mnras, 370, 645

\bibitem[\protect\citeauthoryear{{Bregman}, {Fabian}, {Miller} \&
  {Irwin}}{{Bregman} et~al.}{2006}]{bregman_etal06}
{Bregman} J.~N.,  {Fabian} A.~C.,  {Miller} E.~D.,    {Irwin} J.~A.,  2006,
  \apj, 642, 746

\bibitem[\protect\citeauthoryear{{Brighenti} \& {Mathews}}{{Brighenti} \&
  {Mathews}}{2006}]{brighenti_mathews06}
{Brighenti} F.,  {Mathews} W.~G.,  2006, \apj, 643, 120

\bibitem[\protect\citeauthoryear{{Bullock}, {Kolatt}, {Sigad}, {Somerville},
  {Kravtsov}, {Klypin}, {Primack} \& {Dekel}}{{Bullock}
  et~al.}{2001}]{bullock_etal01}
{Bullock} J.~S.,  {Kolatt} T.~S.,  {Sigad} Y.,  {Somerville} R.~S.,  {Kravtsov}
  A.~V.,  {Klypin} A.~A.,  {Primack} J.~R.,    {Dekel} A.,  2001, \mnras, 321,
  559

\bibitem[\protect\citeauthoryear{{Carilli}, {Perley} \& {Harris}}{{Carilli}
  et~al.}{1994}]{carilli_etal94}
{Carilli} C.~L.,  {Perley} R.~A.,    {Harris} D.~E.,  1994, \mnras, 270, 173

\bibitem[\protect\citeauthoryear{{Cattaneo}, {Blaizot}, {Weinberg}, {Colombi},
  {Dave}, {Devriendt}, {Guiderdoni}, {Katz} \& {Keres}}{{Cattaneo}
  et~al.}{2007}]{cattaneo_etal07}
{Cattaneo} A.,  {Blaizot} J.,  {Weinberg} D.~H.,  {Colombi} S.,  {Dave} R.,
  {Devriendt} J.,  {Guiderdoni} B.,  {Katz} N.,    {Keres} D.,  2007, ArXiv
  Astrophysics e-prints

\bibitem[\protect\citeauthoryear{{Cattaneo}, {Dekel}, {Devriendt}, {Guiderdoni}
  \& {Blaizot}}{{Cattaneo} et~al.}{2006}]{cattaneo_etal06}
{Cattaneo} A.,  {Dekel} A.,  {Devriendt} J.,  {Guiderdoni} B.,    {Blaizot} J.,
   2006, \mnras, 370, 1651

\bibitem[\protect\citeauthoryear{{Churazov}, {Br{\" u}ggen}, {Kaiser}, {B{\"
  o}hringer} \& {Forman}}{{Churazov} et~al.}{2001}]{churazov_etal01}
{Churazov} E.,  {Br{\" u}ggen} M.,  {Kaiser} C.~R.,  {B{\" o}hringer} H.,
  {Forman} W.,  2001, \apj, 554, 261

\bibitem[\protect\citeauthoryear{{Ciotti} \& {Ostriker}}{{Ciotti} \&
  {Ostriker}}{1997}]{ciotti_ostriker97}
{Ciotti} L.,  {Ostriker} J.~P.,  1997, \apjl, 487, L105

\bibitem[\protect\citeauthoryear{{Ciotti} \& {Ostriker}}{{Ciotti} \&
  {Ostriker}}{2001}]{ciotti_ostriker01}
{Ciotti} L.,  {Ostriker} J.~P.,  2001, \apj, 551, 131

\bibitem[\protect\citeauthoryear{{Croton}, {Springel}, {White}, {De Lucia},
  {Frenk}, {Gao}, {Jenkins}, {Kauffmann}, {Navarro} \& {Yoshida}}{{Croton}
  et~al.}{2006}]{croton_etal06}
{Croton} D.~J.,  {Springel} V.,  {White} S.~D.~M.,  {De Lucia} G.,  {Frenk}
  C.~S.,  {Gao} L.,  {Jenkins} A.,  {Kauffmann} G.,  {Navarro} J.~F.,
  {Yoshida} N.,  2006, \mnras, 365, 11

\bibitem[\protect\citeauthoryear{{David}, {Nulsen}, {McNamara}, {Forman},
  {Jones}, {Ponman}, {Robertson} \& {Wise}}{{David}
  et~al.}{2001}]{david_etal01}
{David} L.~P.,  {Nulsen} P.~E.~J.,  {McNamara} B.~R.,  {Forman} W.,  {Jones}
  C.,  {Ponman} T.,  {Robertson} B.,    {Wise} M.,  2001, \apj, 557, 546

\bibitem[\protect\citeauthoryear{{Di Matteo}, {Allen}, {Fabian}, {Wilson} \&
  {Young}}{{Di Matteo} et~al.}{2003}]{dimatteo_etal03}
{Di Matteo} T.,  {Allen} S.~W.,  {Fabian} A.~C.,  {Wilson} A.~S.,    {Young}
  A.~J.,  2003, \apj, 582, 133

\bibitem[\protect\citeauthoryear{{Fabian}}{{Fabian}}{1994}]{fabian94}
{Fabian} A.~C.,  1994, \ARAA, 32, 277

\bibitem[\protect\citeauthoryear{{Fabian}, {Celotti}, {Blundell}, {Kassim} \&
  {Perley}}{{Fabian} et~al.}{2002}]{fabian_etal02}
{Fabian} A.~C.,  {Celotti} A.,  {Blundell} K.~M.,  {Kassim} N.~E.,    {Perley}
  R.~A.,  2002, \mnras, 331, 369

\bibitem[\protect\citeauthoryear{{Fabian}, {Sanders}, {Allen}, {Crawford},
  {Iwasawa}, {Johnstone}, {Schmidt} \& {Taylor}}{{Fabian}
  et~al.}{2003}]{fabian_etal03}
{Fabian} A.~C.,  {Sanders} J.~S.,  {Allen} S.~W.,  {Crawford} C.~S.,  {Iwasawa}
  K.,  {Johnstone} R.~M.,  {Schmidt} R.~W.,    {Taylor} G.~B.,  2003, \mnras,
  344, L43

\bibitem[\protect\citeauthoryear{{Fabian}, {Sanders}, {Ettori}, {Taylor},
  {Allen}, {Crawford}, {Iwasawa}, {Johnstone} \& {Ogle}}{{Fabian}
  et~al.}{2000}]{fabian_etal00}
{Fabian} A.~C.,  {Sanders} J.~S.,  {Ettori} S.,  {Taylor} G.~B.,  {Allen}
  S.~W.,  {Crawford} C.~S.,  {Iwasawa} K.,  {Johnstone} R.~M.,    {Ogle} P.~M.,
   2000, \mnras, 318, L65

\bibitem[\protect\citeauthoryear{{Fabian}, {Sanders}, {Taylor}, {Allen},
  {Crawford}, {Johnstone} \& {Iwasawa}}{{Fabian} et~al.}{2006}]{fabian_etal06}
{Fabian} A.~C.,  {Sanders} J.~S.,  {Taylor} G.~B.,  {Allen} S.~W.,  {Crawford}
  C.~S.,  {Johnstone} R.~M.,    {Iwasawa} K.,  2006, \mnras, 366, 417

\bibitem[\protect\citeauthoryear{{Komatsu} \& {Seljak}}{{Komatsu} \&
  {Seljak}}{2001}]{komatsu_seljak01}
{Komatsu} E.,  {Seljak} U.,  2001, \mnras, 327, 1353

\bibitem[\protect\citeauthoryear{{Mamon} \& {{\L}okas}}{{Mamon} \&
  {{\L}okas}}{2005}]{mamon_lokas05}
{Mamon} G.~A.,  {{\L}okas} E.~L.,  2005, \mnras, 362, 95

\bibitem[\protect\citeauthoryear{{Mathews}, {Faltenbacher} \&
  {Brighenti}}{{Mathews} et~al.}{2006}]{mathews_etal06}
{Mathews} W.~G.,  {Faltenbacher} A.,    {Brighenti} F.,  2006, \apj, 638, 659

\bibitem[\protect\citeauthoryear{{McCarthy}, {Babul} \& {Balogh}}{{McCarthy}
  et~al.}{2002}]{mccarthy_etal02}
{McCarthy} I.~G.,  {Babul} A.,    {Balogh} M.~L.,  2002, \apj, 573, 515

\bibitem[\protect\citeauthoryear{{McCarthy}, {Balogh}, {Babul}, {Poole} \&
  {Horner}}{{McCarthy} et~al.}{2004}]{mccarthy_etal04}
{McCarthy} I.~G.,  {Balogh} M.~L.,  {Babul} A.,  {Poole} G.~B.,    {Horner}
  D.~J.,  2004, \apj, 613, 811

\bibitem[\protect\citeauthoryear{{McNamara}, {Nulsen}, {Wise}, {Rafferty},
  {Carilli}, {Sarazin} \& {Blanton}}{{McNamara} et~al.}{2005}]{mcnamara_etal05}
{McNamara} B.~R.,  {Nulsen} P.~E.~J.,  {Wise} M.~W.,  {Rafferty} D.~A.,
  {Carilli} C.,  {Sarazin} C.~L.,    {Blanton} E.~L.,  2005, \nat, 433, 45

\bibitem[\protect\citeauthoryear{{McNamara}, {Wise}, {Nulsen}, {David},
  {Sarazin}, {Bautz}, {Markevitch}, {Vikhlinin}, {Forman}, {Jones} \&
  {Harris}}{{McNamara} et~al.}{2000}]{mcnamara_etal00}
{McNamara} B.~R.,  {Wise} M.,  {Nulsen} P.~E.~J.,  {David} L.~P.,  {Sarazin}
  C.~L.,  {Bautz} M.,  {Markevitch} M.,  {Vikhlinin} A.,  {Forman} W.~R.,
  {Jones} C.,    {Harris} D.~E.,  2000, \apjl, 534, L135

\bibitem[\protect\citeauthoryear{{McNamara}, {Wise}, {Nulsen}, {David},
  {Carilli}, {Sarazin}, {O'Dea}, {Houck}, {Donahue}, {Baum}, {Voit},
  {O'Connell} \& {Koekemoer}}{{McNamara} et~al.}{2001}]{mcnamara_etal01}
{McNamara} B.~R.,  {Wise} M.~W.,  {Nulsen} P.~E.~J.,  {David} L.~P.,  {Carilli}
  C.~L.,  {Sarazin} C.~L.,  {O'Dea} C.~P.,  {Houck} J.,  {Donahue} M.,  {Baum}
  S.,  {Voit} M.,  {O'Connell} R.~W.,    {Koekemoer} A.,  2001, \apjl, 562,
  L149

\bibitem[\protect\citeauthoryear{{Navarro}, {Frenk} \& {White}}{{Navarro}
  et~al.}{1996}]{navarro_etal96}
{Navarro} J.~F.,  {Frenk} C.~S.,    {White} S.~D.~M.,  1996, \apj, 462, 563

\bibitem[\protect\citeauthoryear{{Navarro}, {Frenk} \& {White}}{{Navarro}
  et~al.}{1997}]{navarro_etal97}
{Navarro} J.~F.,  {Frenk} C.~S.,    {White} S.~D.~M.,  1997, \apj, 490, 493

\bibitem[\protect\citeauthoryear{{Nusser}, {Silk} \& {Babul}}{{Nusser}
  et~al.}{2006}]{nusser_etal06}
{Nusser} A.,  {Silk} J.,    {Babul} A.,  2006, ArXiv Astrophysics e-prints

\bibitem[\protect\citeauthoryear{{Oh} \& {Benson}}{{Oh} \&
  {Benson}}{2003}]{oh_benson03}
{Oh} S.~P.,  {Benson} A.~J.,  2003, \mnras, 342, 664

\bibitem[\protect\citeauthoryear{{Omma} \& {Binney}}{{Omma} \&
  {Binney}}{2004}]{omma_binney04}
{Omma} H.,  {Binney} J.,  2004, \mnras, 350, L13

\bibitem[\protect\citeauthoryear{{Omma}, {Binney}, {Bryan} \& {Slyz}}{{Omma}
  et~al.}{2004}]{omma_etal04}
{Omma} H.,  {Binney} J.,  {Bryan} G.,    {Slyz} A.,  2004, \mnras, 348, 1105

\bibitem[\protect\citeauthoryear{{Owen}, {Eilek} \& {Kassim}}{{Owen}
  et~al.}{2000}]{owen_etal00}
{Owen} F.~N.,  {Eilek} J.~A.,    {Kassim} N.~E.,  2000, \apj, 543, 611

\bibitem[\protect\citeauthoryear{{Peterson}, {Kahn}, {Paerels}, {Kaastra},
  {Tamura}, {Bleeker}, {Ferrigno} \& {Jernigan}}{{Peterson}
  et~al.}{2003}]{peterson_etal03}
{Peterson} J.~R.,  {Kahn} S.~M.,  {Paerels} F.~B.~S.,  {Kaastra} J.~S.,
  {Tamura} T.,  {Bleeker} J.~A.~M.,  {Ferrigno} C.,    {Jernigan} J.~G.,  2003,
  \apj, 590, 207

\bibitem[\protect\citeauthoryear{{Quilis}, {Bower} \& {Balogh}}{{Quilis}
  et~al.}{2001}]{quilis_etal01}
{Quilis} V.,  {Bower} R.~G.,    {Balogh} M.~L.,  2001, \mnras, 328, 1091

\bibitem[\protect\citeauthoryear{{Reynolds}, {Heinz} \& {Begelman}}{{Reynolds}
  et~al.}{2001}]{reynolds_etal01}
{Reynolds} C.~S.,  {Heinz} S.,    {Begelman} M.~C.,  2001, \apjl, 549, L179

\bibitem[\protect\citeauthoryear{{Reynolds}, {Heinz} \& {Begelman}}{{Reynolds}
  et~al.}{2002}]{reynolds_etal02}
{Reynolds} C.~S.,  {Heinz} S.,    {Begelman} M.~C.,  2002, \mnras, 332, 271

\bibitem[\protect\citeauthoryear{{Roychowdhury}, {Ruszkowski}, {Nath} \&
  {Begelman}}{{Roychowdhury} et~al.}{2004}]{roy_etal04}
{Roychowdhury} S.,  {Ruszkowski} M.,  {Nath} B.~B.,    {Begelman} M.~C.,  2004,
  \apj, 615, 681

\bibitem[\protect\citeauthoryear{{Ruszkowski}, {Br{\"u}ggen} \&
  {Begelman}}{{Ruszkowski} et~al.}{2004}]{ruszkowski_etal04}
{Ruszkowski} M.,  {Br{\"u}ggen} M.,    {Begelman} M.~C.,  2004, \apj, 615, 675

\bibitem[\protect\citeauthoryear{{Salom{\'e}}, {Combes}, {Edge}, {Crawford},
  {Erlund}, {Fabian}, {Hatch}, {Johnstone}, {Sanders} \& {Wilman}}{{Salom{\'e}}
  et~al.}{2006}]{salome_etal06}
{Salom{\'e}} P.,  {Combes} F.,  {Edge} A.~C.,  {Crawford} C.,  {Erlund} M.,
  {Fabian} A.~C.,  {Hatch} N.~A.,  {Johnstone} R.~M.,  {Sanders} J.~S.,
  {Wilman} R.~J.,  2006, \aap, 454, 437

\bibitem[\protect\citeauthoryear{{Sazonov}, {Ostriker} \& {Sunyaev}}{{Sazonov}
  et~al.}{2004}]{sazonov_etal04}
{Sazonov} S.~Y.,  {Ostriker} J.~P.,    {Sunyaev} R.~A.,  2004, \mnras, 347, 144

\bibitem[\protect\citeauthoryear{{Seljak}}{{Seljak}}{2000}]{seljak00}
{Seljak} U.,  2000, \mnras, 318, 203

\bibitem[\protect\citeauthoryear{{Sijacki} \& {Springel}}{{Sijacki} \&
  {Springel}}{2006}]{sijacki_springel06}
{Sijacki} D.,  {Springel} V.,  2006, \mnras, 366, 397

\bibitem[\protect\citeauthoryear{{Smith}, {Smarr}, {Norman} \&
  {Wilson}}{{Smith} et~al.}{1983}]{smith_etal83}
{Smith} M.~D.,  {Smarr} L.,  {Norman} M.~L.,    {Wilson} J.~R.,  1983, \apj,
  264, 432

\bibitem[\protect\citeauthoryear{{Sutherland} \& {Dopita}}{{Sutherland} \&
  {Dopita}}{1993}]{sutherland_dopita93}
{Sutherland} R.~S.,  {Dopita} M.~A.,  1993, \apjs, 88, 253

\bibitem[\protect\citeauthoryear{{Tabor} \& {Binney}}{{Tabor} \&
  {Binney}}{1993}]{tabor_binney93}
{Tabor} G.,  {Binney} J.,  1993, \mnras, 263, 323

\bibitem[\protect\citeauthoryear{{Teyssier}}{{Teyssier}}{2002}]{teyssier02}
{Teyssier} R.,  2002, \aap, 385, 337

\bibitem[\protect\citeauthoryear{{Tucker} \& {David}}{{Tucker} \&
  {David}}{1997}]{tucker_david97}
{Tucker} W.,  {David} L.~P.,  1997, \apj, 484, 602

\bibitem[\protect\citeauthoryear{{Vernaleo} \& {Reynolds}}{{Vernaleo} \&
  {Reynolds}}{2006}]{vernaleo_reynolds06}
{Vernaleo} J.~C.,  {Reynolds} C.~S.,  2006, \apj, 645, 83

\bibitem[\protect\citeauthoryear{{Voigt} \& {Fabian}}{{Voigt} \&
  {Fabian}}{2004}]{voigt_fabian04}
{Voigt} L.~M.,  {Fabian} A.~C.,  2004, \mnras, 347, 1130

\bibitem[\protect\citeauthoryear{{Yu} \& {Tremaine}}{{Yu} \&
  {Tremaine}}{2002}]{yu_tremaine02}
{Yu} Q.,  {Tremaine} S.,  2002, \mnras, 335, 965

\bibitem[\protect\citeauthoryear{{Zanni}, {Murante}, {Bodo}, {Massaglia},
  {Rossi} \& {Ferrari}}{{Zanni} et~al.}{2005}]{zanni_etal05}
{Zanni} C.,  {Murante} G.,  {Bodo} G.,  {Massaglia} S.,  {Rossi} P.,
  {Ferrari} A.,  2005, \aap, 429, 399

\end{thebibliography}

\appendix

\section{The Komatsu \& Seljak hydrostatic solution for the hot gas of an NFW halo}

If the distribution of the dark matter is governed by the NFW \citep{navarro_etal96,navarro_etal97}
density profile, then the equation of hydrostatic equilibrium reads: 
\begin{equation}
\label{hydrostatic}
{1\over\rho}{{\rm d}p\over{\rm d}r}
=-{{\rm G}M_{\rm vir}\over r^2}
{ {\rm ln}(1+cx)-c x/(1+c x) \over {\rm ln}(1+c)-c/(1+c)}.
\end{equation}
Here $M_{\rm vir}$ is the halo mass within the virial radius $r_{\rm vir}$ and $ x\equiv r/r_{\rm vir}$.
For the NFW concentration parameter, $c$, we follow \citet{seljak00} and adopt: 
\begin{equation}
\label{cnfw}
c=6(M_{\rm vir}/10^{14}M_\odot)^{-0.2}.
\end{equation}
\citet{bullock_etal01} have argued that a model in which the concentration depends on the redshift of collapse
agrees better with the observation, but Eq.~(\ref{cnfw}) is perfectly adequate for the purpose of constructing a density
profile that approximates that of the ICM of a galaxy cluster.

To be able to solve Eq.~(\ref{hydrostatic}) we must introduce an additional constraint on the temperature profile.
An isothermal sphere is inconsistent with the requirement that the gas traces the dark matter in the outer
parts of the halo. Instead, \citet{komatsu_seljak01} find a good agreement
with the measured X-ray profiles and with the 
observed mass-temperature relation when the gas is described as a polytrope, with the relation
\begin{equation}
\label{polytropic}
p=(\gamma_{\rm p}-1)\rho_0\epsilon_0(\rho/\rho_0)^{\gamma_{\rm p}}.
\end{equation}
Here $\rho_0$ and $\epsilon_0$ are the density and the specific internal energy of the gas at the bottom of the potential
well, while $\gamma_{\rm p}$ is the polytropic index. 
By substituting Eq.~(\ref{polytropic}) into Eq.~(\ref{hydrostatic}), we find that:
\begin{equation}
\label{hydrostatic2}
\gamma_{\rm p}\left({\rho\over\rho_0}\right)^{\gamma_{\rm p}-1}{{\rm d}\over{\rm d} x}{\rho\over\rho_0}=
-{1\over\eta_0}{1\over x^2}{ {\rm ln}(1+c x)-c x/(1+c x) \over {\rm ln}(1+c)-c/(1+c)},
\end{equation}
where 
\begin{equation}
\eta_0^{-1}\equiv{{\rm G}M_{\rm vir}\over r_{\rm vir}(\gamma_{\rm p}-1)\epsilon_0}=
{{\rm G}\mu M_{\rm vir}\over r_{\rm vir}kT_0}
\end{equation}
and $T_0$ is the gas temperature in the cluster core. The general solution of Eq.~(\ref{hydrostatic2}) is
\begin{equation}
{\rho\over\rho_0}=\left\{1-{\gamma_{\rm p}-1\over\gamma_{\rm p}\eta_0}
{c\over{\rm ln}(1+c)-{c\over 1+c}}\left[1-{{\rm ln}(1+c x)\over c x}\right]\right\}^{1\over\gamma_{\rm p}-1}
\end{equation}
where the parameters $\rho_0$ and $\eta_0$ are determined by imposing 
that $\rho$ satisfies the conditions:
\begin{equation}
\label{densnorm}
\rho(r_{\rm vir})=f_{\rm b}\rho_{\rm dm}(r_{\rm vir})
\end{equation}
and
\begin{equation}
\label{slope}
{{\rm d\,ln}\rho\over{\rm d\,ln}r}(r_{\rm vir})={{\rm d\,ln}\rho_{\rm dm}\over{\rm d\,ln}r}(r_{\rm vir}).
\end{equation}
Here $\rho_{\rm dm}$ is the dark matter density distribution in the NFW model and $f_{\rm b}=0.1$ is the baryonic mass 
fraction.
Eq.~(\ref{densnorm}) gives:
\begin{equation}
\rho_0={\rho_0\over\rho(r_{\rm vir})}\cdot {f_{\rm b}\rho_{\rm dm}(r_{\rm vir})\over\rho_{\rm vir}}
\cdot{3M_{\rm vir}\over 4\pi r_{\rm vir}^3}=
\end{equation}
$$={\rho_0\over\rho}( x=1)\cdot{c^2\over 3(1+c)^2}{1\over{\rm ln}(1+c)-c/(1+c)}\rho_{\rm vir}.$$
q.~(\ref{slope}) gives:
\begin{equation}
\eta_0={1\over\gamma_{\rm p}}\left[{c+1\over 3c+1}+(\gamma_{\rm p}-1){c-{\rm ln}(1+c)\over{\rm ln}(1+c)-c/(1+c)}\right].
\end{equation}
The conditions (\ref{densnorm}) and (\ref{slope}) guarantee
that the gas density profile and the dark matter density profile have
the appropriate normalization and have the same slope at $r\simeq r_{\rm vir}$, 
but do not guarantee that the two profile will not diverge at  $r> r_{\rm vir}$.
However, \citet{komatsu_seljak01} point out that one can fine tune the
polytropic index $\gamma_{\rm p}$ to
\begin{equation}
\gamma_{\rm p}=1.15+0.01(c-6.5)
\end{equation}
so that $\rho\simeq f_{\rm b}\rho_{\rm dm}$ for
$r_{\rm vir}/2<r<2r_{\rm vir}$.

\end{document}